\documentclass[twocolumn,superscriptaddress,showpacs,longbibliography, aps, prb]{revtex4-1}
\usepackage[dvipdfmx]{graphicx}
\usepackage{epstopdf}
\usepackage{hyperref}
\usepackage{dcolumn}
\usepackage{bm}
\usepackage{natbib}
\usepackage{xspace}
\usepackage{color}
\usepackage{ulem}
\usepackage{amsmath}
\usepackage{amssymb}
\hypersetup{colorlinks=true,breaklinks,linkcolor=blue,urlcolor=blue,citecolor=blue}
\begin{document}
\newcommand{\bR}{\mbox{\boldmath $R$}}
\newcommand{\trs}[1]{\textcolor{red}{\sout{#1}}}
\newcommand{\tb}[1]{\textcolor{blue}{#1}}
\newcommand{\tbs}[1]{\textcolor{blue}{\sout{#1}}}
\newcommand{\tc}[1]{\textcolor{cyan}{#1}}
\newcommand{\tcs}[1]{\textcolor{cyan}{\sout{#1}}}
\newcommand{\Ha}{\mathcal{H}}
\newcommand{\mh}{\mathsf{h}}
\newcommand{\mA}{\mathsf{A}}
\newcommand{\mB}{\mathsf{B}}
\newcommand{\mC}{\mathsf{C}}
\newcommand{\mS}{\mathsf{S}}
\newcommand{\mU}{\mathsf{U}}
\newcommand{\mX}{\mathsf{X}}
\newcommand{\sP}{\mathcal{P}}
\newcommand{\sL}{\mathcal{L}}
\newcommand{\sO}{\mathcal{O}}
\newcommand{\la}{\langle}
\newcommand{\ra}{\rangle}
\newcommand{\ga}{\alpha}
\newcommand{\gb}{\beta}
\newcommand{\gc}{\gamma}
\newcommand{\gs}{\sigma}
\newcommand{\vd}{{\bm{d}}}
\newcommand{\vb}{{\bm{b}}}
\newcommand{\vc}{{\bm{c}}}
\newcommand{\vs}{{\bm{s}}}
\newcommand{\vk}{{\bm{k}}}
\newcommand{\vq}{{\bm{q}}}
\newcommand{\vR}{{\bm{R}}}
\newcommand{\vQ}{{\bm{Q}}}
\newcommand{\vga}{{\bm{\alpha}}}
\newcommand{\vgc}{{\bm{\gamma}}}
\newcommand{\mb}[1]{\mathbf{#1}}
\arraycolsep=0.0em
\newcommand{\Ns}{N_{\text{s}}}
\def\vec#1{\boldsymbol #1}

\title{
Zeros of Green Functions in Topological Insulators
}

\author{Takahiro Misawa}
\affiliation{Beijing Academy of Quantum Information Sciences, Haidian District, Beijing 100193, China}
\affiliation{Research Institute for Science and Engineering, Waseda University, 3-4-1, Okubo, Shinjuku, Tokyo 169–8555, Japan}
\affiliation{Institute for Solid State Physics,~University of Tokyo,~5-1-5 Kashiwanoha, Kashiwa, Chiba 277-8581, Japan}
\author{Youhei Yamaji}
\affiliation{
Center for Green Research on Energy and Environmental Materials, National Institute for Materials Science, Namiki, Tsukuba-shi, Ibaraki, 305-0044, Japan}
\affiliation{Department of Applied Physics, University of Tokyo, 7-3-1 Hongo, Bunkyo-ku, Tokyo 113–8656, Japan}

\date{\today}
\begin{abstract}
This study demonstrates that the zeros of the diagonal components of 
Green functions are key quantities 
that can detect {non-interacting topological insulators}. 
{We} {show} that 
{zeros of the Green functions traverse the band gap in the topological phases.
The traverses induce the crosses {of zeros,} and the zeros' surface in the band gap, {analogous to the Fermi surface
of metals}.}
{By calculating the zeros of the microscopic models, we show the traverses of the zeros}
universally appear in all six classes of conventional {non-interacting topological insulators}.
{By utilizing the eigenvector-eigenvalue identity, which is a recently rediscovered relation in linear algebra,
we prove}
that {the traverses} of the zeros in the bulk Green functions 
are guaranteed {by the band inversions, which occur in the topological phases.}
The relevance of the zeros {to} detecting 
the exotic topological insulators such 
as the higher-order topological insulators is also discussed.
{For the Hamiltonians with the nearest-neighbor hoppings,
we also show that the gapless edge state guarantees 
the zeros' surfaces in the band gap.}
The analysis demonstrates that the zeros can be used to 
{detect a wide range of topological insulators}
and thus useful for searching {new} topological materials.
\end{abstract}

\maketitle

\section{Introduction}
The Green function is one of the most
fundamental tools used for solving
quantum many-body problems~\cite{AGD,Fetter}. 
The Green function method can be used to perform a systematic analysis for
a wide range of quantum many-body systems, 
from elementary particle physics to condensed matter physics.
One of the key components of the Green function is the pole, 
at which the value of the Green function becomes infinite.
In solid state physics, the poles correspond to
band dispersion in solids and play a central role in describing the
low-energy excitations of solids. 

{In insulating phases,}
the zeros of the Green function, rather than the poles, appear in the bandgap. The zeros 
can be used for characterizing the insulating phases, 
wherein the poles (band dispersions) do not govern low-energy physics.
In fact, it has been proposed that the surface of the zeros of Green functions (zeros' surfaces) in
Mott insulators can act as Fermi surfaces in metallic phases~\cite{Dzyaloshinskii}.
For particle--hole symmetric systems,
it has been demonstrated that the Luttinger's 
sum rules on the zeros' surface 
can be satisfied for the insulating phases{~\cite{Stanescu_PRB2007,Seki_PRB2017}.}
Moreover, studies on the doped Mott insulators
have established that the interplay of the zeros and poles of the Green functions can govern
the unconventional electronic properties 
such as non-Fermi liquid behaviors, pseudo-gap phenomena, 
and Fermi arc structures that are observed in high-$T_{\rm c}$ cuprates
{~\cite{Stanescu_PRB2006,Dave_PRL2013,SakaiPRB,YamajiPRB}.}

Recent findings on topological insulators~\cite{KaneMele}
have triggered intensive experimental and theoretical
investigations of topological materials~\cite{KaneRMP,Qi_RMP2011,Ando_JPSJ2013,Chiu_RMP2016}.
Systematic clarifications of topological insulators have been proposed~\cite{RyuFurusaki,Kitaev_AIP2009,Chiu_RMP2016},
and 
the periodic table for six conventional classes of
topological insulators has been established, as presented in Table~\ref{PeriodicTable}.
It has also been demonstrated that
the existence of the topological 
invariant is 
related with the gapless surface states~\cite{Essin_PRB2011}. 

Although topological insulators can be detected by calculating
the topological invariants,
direct calculations of the topological 
invariants are generally not easy.
Thus, several simplified methods have been proposed 
for detecting the topological phases, such as the Fu-Kane formula~\cite{Fu_RPB2007}
for the Z$_2$ topological insulators with {the} inversion symmetry. 
{Recently, classification methods utilizing the symmetries of solids
such as the symmetry eigenvalues~\cite{Kruthoff_PRX2017} and
the symmetry indicators\cite{Po_NCom2017} 
were proposed.
In particular, the symmetry indicators
have been used for searching
{a wide range of
materials
for topologically non-trivial phases}~\cite{Bradlyn_Nature2017,Tang_NPhys2019,Zhang_Nature2019,Tang_Nature2019,Vergniory_Nature2019,Tang_SA2019,Xu_2020Nature,Iraola_PRB2021}.}

\begin{table}[tb]
\begin{tabular}{lcccrrrrr} \hline
class   & ~~$\Theta$ & ~$\Xi$ & ~$\Gamma$ & ~~~$d=1$   & ~~~2   & ~~~3    & ~~~4   \\ \hline
A       & 0        & 0     & 0            & 0     & $\mathbb{Z}$     & 0      & $\mathbb{Z}$     \\
AI      & 1        & 0     & 0            & 0     & 0     & 0      & 2$\mathbb{Z}$    \\
AII     & -1       & 0     & 0            & 0     & $\mathbb{Z}_2$ & $\mathbb{Z}_2$  & $\mathbb{Z}$     \\ \hline
AIII    & 0        & 0     & 1            & $\mathbb{Z}$     & 0     & $\mathbb{Z}$      & 0     \\
BDI     & -1       & -1    & 1            & $\mathbb{Z}$     & 0     & 0      & 0     \\ 
CII     & 1        & 1     & 1            & 2$\mathbb{Z}$    & 0     & $\mathbb{Z}_2$  & $\mathbb{Z}_2$ \\ \hline
\end{tabular}
\caption{Periodic table of topological insulators under four dimensions~\cite{RyuFurusaki}.
The indices $\Theta$, $\Xi$, and $\Gamma$ represent
the time-reversal symmetry, particle--hole symmetry,
and chiral symmetry.
For topological insulators without 
chiral symmetry ($\Gamma=0$, upper panel),
this study demonstrates that the zeros of the Green function 
{traverse} the bandgap.
For topological insulators with 
chiral symmetry ($\Gamma=1$, lower panel),
this investigation establishes that it is necessary to perform an appropriate unitary 
transformation to observe the {traverse} of the zeros.
}
\label{PeriodicTable}
\end{table}

{The present} study shows that the zeros of the {diagonal components of the} Green functions
are useful quantities for detecting the {non-interacting} topological insulators.
Because the microscopic Hamiltonians for the topological
insulators {in non-interacting systems} are generally multi-orbital systems,
we should consider the matrices of the Green functions.
This investigation focuses on the zeros of 
the diagonal components of the Green functions
because they have characteristic features.
These zeros exist between poles and are given by the 
eigenvalues of the minor matrix $M_{n}$, which is obtained
by removing the $n$th row and column from the original Hamiltonian.
It is noted that the simple and 
fundamental relationship between the zeros of the Green functions 
and the minor matrix is not well known.
This study proves that these characteristic features are 
{obtained from}
an argument that is provided in Ref.~[\onlinecite{Denton_BAMS2022}], 
which rediscovers the fundamental but lesser-known identity between 
the eigenvectors and the eigenvalues called the {\it eigenvector-eigenvalue identity}.
The fact that the eigenvector-eigenvalue identity is not widely known may be 
the main reason why the relationship 
between the zeros and the minor matrix is not well understood.
The relationship with the minor matrix 
offers a mathematical foundation 
that can be used to analyze the 
zeros of {the diagonal components of} the Green functions. 
It also offers an efficient way to numerically 
obtain the zeros of the {diagonal components of the} Green functions, i.e., 
it is {\it not} necessary to search for the zeros in the energy direction.

{{While}
the importance of the zeros of the Green functions 
for characterizing the interacting topological insulators
has been proposed by Gurarie in Ref.~[\onlinecite{Gurarie}],
the definition of the zeros of the Green functions for {the present} study 
is different from the definition used in Gurarie's work. 
In Gurarie's work,
{the zero eigenvalues of the Green functions
(in other words, zeros of the 
diagonalized Green functions)
are defined as zeros of the Green functions}.
Because the {\it diagonalized} Green functions have no zeros 
in the non-interacting case,
Gurarie's method cannot be applied to 
characterize non-interacting topological insulators.
In contrast, our argument is based on the 
diagonal components of the Green functions that can 
characterize the non-interacting topological insulators,
as will be described in detail below.
It is noted that the diagonalized 
Green function at the zero frequency~\cite{Wang_RPX2012,Wang_IOP2013}
is used for detecting the interacting topological phases.
{We also note {that} the zeros of the diagonal component of the Green functions
were used for detecting impurity states~\cite{Gonzalez_PRB2012,Slager_PRB2015} 
and the appearance of the Majorana bound states~\cite{Alvarado_PRB2020}.}}

By using the mathematical foundation of the 
zeros for the diagonal component of the Green functions,
{we clarify} their behavior in topological insulators.
First, by taking the two-dimensional 
Chern insulator as an example~(class A topological insulator in Table~\ref{PeriodicTable}), 
it is demonstrated that 
{the zeros in the bulk Green functions traverse the bandgap
in topological insulators; however,
the zeros do not traverse in trivial insulators.}
{We establish} 
that the existence of the 
topological invariant (Chern number) guarantees the 
{traverse} of the zeros in the bulk systems. 
{As the consequence of
{the traverses of the zeros from the different diagonal components of the Green functions},
we find that the crosses of the zeros appear in the topological phases.}
{We give a proof that the band inversion, which generally occurs in topological phases, 
guarantees the traverse of the zeros in the topological phases in any dimensions.}

It is also demonstrated that the 
{traverse} of the zeros generically occurs in the
other topological insulators such
as the $\mathbb{Z}_2$ topological
insulator in two and three dimensions 
(class AII topological insulators in Table~\ref{PeriodicTable}) as 
well as in $2\mathbb{Z}$ topological insulators in four dimensions
(class AI topological insulators in Table~\ref{PeriodicTable}).
This study also investigates the zeros in 
topological insulators 
with chiral symmetry (lower panel in Table~\ref{PeriodicTable}) and shows
that 
{traverses} of the zeros 
also occur in chiral topological insulators when we take 
appropriate gauges. 
{Furthermore, for the Hamiltonian with the nearest neighbor
hoppings, it is also demonstrated that the gapless edge states
induce the zero's surface in the bulk systems.
In other words, at least one diagonal component of the 
Green functions becomes zero in the band gap due to the
gapless edge states.}

Recent studies have suggested the existence of exotic topological phases,
which are not listed in Table.~I. 
Higher-order topological 
insulators~\cite{Benalcazar_Science2017,Schindler_SA2018}, 
which show hinge states or quantized corner charges, 
are candidates for these exotic topological phases.
This study demonstrates that the zeros of Green functions
are also useful for detecting higher-order topological insulators.
These results indicate that {the traverses and the resultant crosses} of the zeros are
universal features of the topological phases.

{Here, we summarize the main features of the zeros of the diagonal components of the 
Green functions that are clarified in this paper:
\begin{enumerate}
\item Zeros of the diagonal component of Green functions are given by the eigenvalues 
of the minor matrices of the original Hamiltonians (Sec. II).
\item Band inversions defined in Eq.~(\ref{eq:bandinversion}) 
induce the traverses of the zeros in the band gap (Sec. V).
\item Since the existence of the Chern number and the $Z_{2}$ topological
invariant guarantees the existence of the band inversions (Appendix~\ref{ap:bandinversion}),
the traverses of the zeros occur in these topological insulators 
under the proper unitary transformation.
\item We show a general way for constructing the 
proper unitary transformation in Eq.~(\ref{eq:unitary}), which induces the
traverses of the zeros in the topological phases. 
This result demonstrates that the traverses of the
zeros are gauge invariant since we can always eliminate 
the apparent gauge dependence with a concrete procedure.
\item Gapless edge states (degeneracy in the edge states) guarantee 
that at least one diagonal component of the Green functions becomes zero in the band gap 
for the Hamiltonians only contain nearest-neighbor hoppings (Sec.~VIII).
\end{enumerate}
}

This paper is organized as follows.~In Sec. II, the mathematical 
foundation of the zeros of the diagonal components of the Green functions is explained. 
In Sec. III, we examine the zeros of the Green function for two-dimensional Chern insulators 
and demonstrate that the {traverse} of the zeroes is induced by the non-trivial Chern number.
We also demonstrate that the {traverse} of the zeroes is gauge invariant. 
 
~Sec. IV establishes that the {traverses} of the zeros occur
in $\mathbb{Z}_2$ topological insulators in two and three dimensions.
{In Sec. V, we give a proof that the traverses 
of the zeros in the topological insulators universally occur due to the band inversions,
which necessarily occur in the topological phases.}
~In Sec. VI, we examine the zeros in 
topological insulators with chiral symmetry
and demonstrate that it is necessary to use appropriate gauges of the Hamiltonian
to see the {traverses} of the zeros in the topological phases.
~Sec. VII shows that the {traverses} of the zeros occur
even for higher-order topological insulators.
{In Sec. VIII, we show that the gapless edge states
guarantee the zeros' surface in the band gap for the
Hamiltonian with the nearest neighbor hoppings.}
Finally, Sec. IX summarizes the study.

\section{Mathematical foundation of zeros of the Green function}
{In this section, we discuss} the mathematical foundation of the 
zeros of the diagonal components of Green functions. 
By following the argument in Ref.~[\onlinecite{Denton_BAMS2022}],
we show that the zeros can be represented by the eigenvalues of the minor
matrix $M_n$ of the Hamiltonian $H$, which is generated by
removing the $n$th row and column from $H$.

Using the eigenvalues and eigenvectors of the Hamiltonian $H(\vec{k})$, where $\vec{k}$ ($\omega$) 
represents the momentum (energy),
the $n$th diagonal component of the Green function is expressed as
\begin{align}
&G_{n}(\vec{k},\omega)
=(\omega I-H(\vec{k}))^{-1}_{nn}
=\sum_{i=1}^{N}\frac{|\Psi_{i}^{(n)}(\vec{k})|^2}{\omega-E_{i}(\vec{k})}, \\
\label{eq:green}
&H(\vec{k})\boldsymbol{\Psi}_{i}(\vec{k})=E_{i}(\vec{k})\boldsymbol{\Psi}_{i}(\vec{k}).
\end{align}
where $E_{i}(\vec{k})$ is the $i$th eigenvalue of the 
Hamiltonian and
$\Psi_{i}^{(n)}(\vec{k})$ is the $n$th component of 
the $i$th eigenvector $\boldsymbol{\Psi}_{i}(\vec{k})$.
By applying {the} Cramer's rule, we can obtain
\begin{align}
&G_{n}(\vec{k},\omega)
=\sum_{i=1}^{N}\frac{|\Psi_{i}^{(n)}(\vec{k})|^2}{\omega-E_{i}(\vec{k})} \notag \\
&=\frac{\det(\omega I_{N-1}-M_{n}(\vec{k}))}{\det(\omega I_{N}-H(\vec{k}))}
=\frac{\prod_{l=1}^{N-1}[\omega-E_{l}(M_{n}(\vec{k}))]}
{\prod_{l=1}^{N}[\omega-E_{l}(H(\vec{k}))]},
\label{eq:minor}
\end{align}
where $I_{N}$ represents the $N$-dimensional identity matrix.
Here, for clarity, we explicitly show the dimensions of the identity matrix.
This relation {Eq.~(\ref{eq:minor})} shows that the 
zeros of $G_{n}(\boldsymbol{k},\omega)$ are represented 
by the eigenvalues of $M_{n}$.
It is known that the eigenvalues of the minor matrix $M_{n}$
exist between the eigenvalues of the original Hamiltonian $H$, i.e.,
\begin{align}
E_{i}(H)\leq E_{i} (M_{n}) \leq E_{i+1}(H).
\end{align}
{The above} relation is known as
the Cauchy interlacing inequalities relation ~\cite{Wilkinson_1963,Hwang_2004}.
A proof of the inequalities is given in Appendix ~\ref{ap:Cauchy}.
{From the definition of the zeros, it is obvious that 
the zeros {form} curves in one dimensions
and surfaces in two dimensions as
{the band dispersions do}.}

By applying the residue theorem to Eq.~(\ref{eq:minor}),
the following expression can be obtained.
\begin{equation}
|\Psi_{i}^{(n)}(\vec{k})|^2=\frac{\prod_{l=1}^{N-1}[E_{i}(H(\vec{k}))-E_{l}(M_{n}(\vec{k}))]}
{\prod_{l=1,l\neq i}^{N}[E_{i}(H(\vec{k}))-E_{l}(H(\vec{k}))]}.
\label{eq:EE}
\end{equation}
This relation {Eq.~(\ref{eq:EE})} is called the {\it eigenvector-eigenvalue identity} and
 has been recently rediscovered~\cite{Denton_BAMS2022}.
From this {identity},
it can be concluded that the zero and the pole
should coincide if the $n$th component of the
eigenvectors becomes zero and vice versa.
{We note that the existence of this special 
point ($|\Psi_{i}^{(n)}(\vec{k})|=0$, called vortex core in the literature~\cite{Kohmoto_1985ANN}) plays
an essential role in detecting the topological phases.}

\begin{figure}[t!]
	\begin{center}
		\includegraphics[width=8cm]{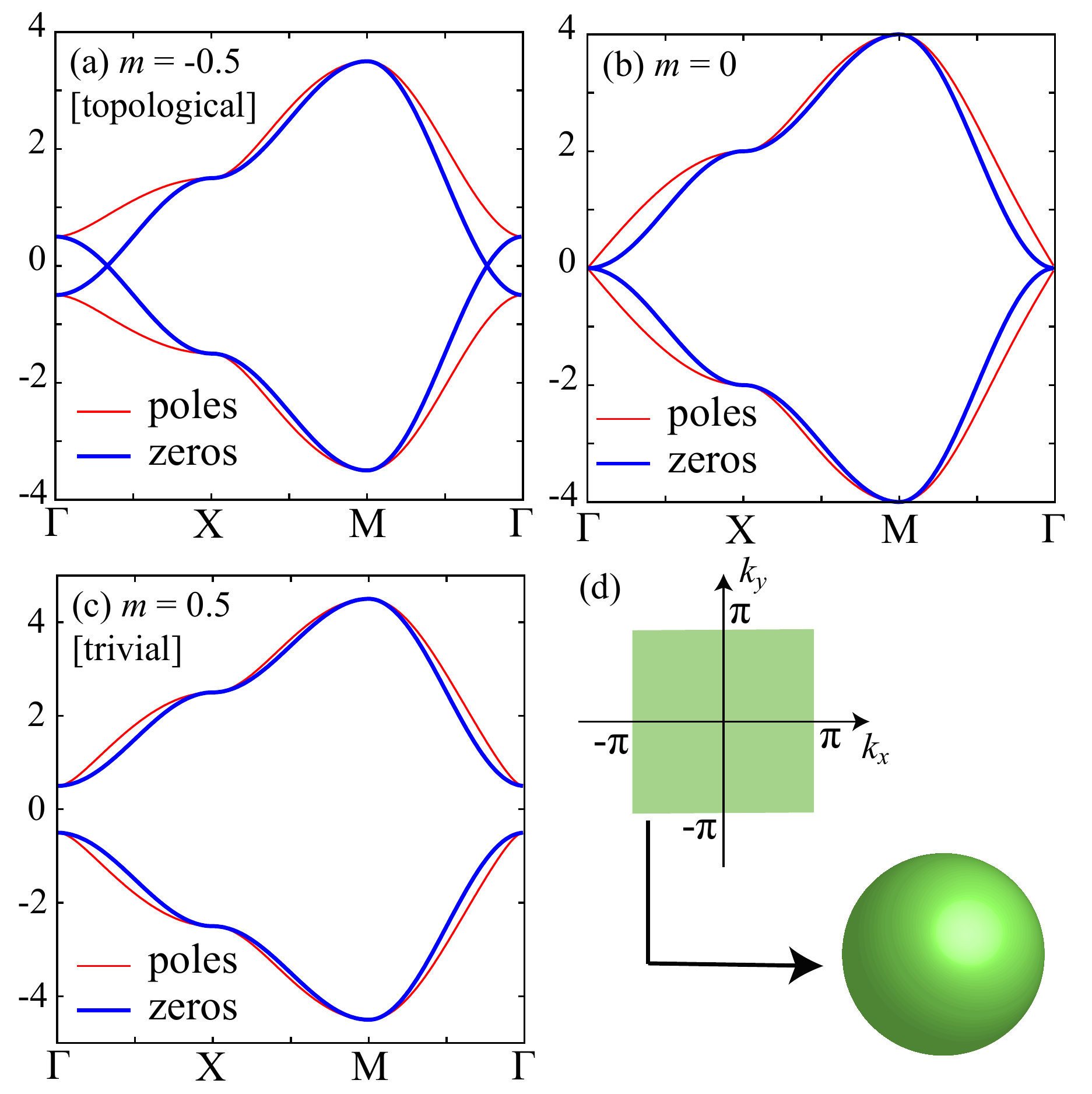}   
	\end{center}
\caption{(a)-(c)~Band dispersions (thin red lines) and 
zeros (thick blue lines) of the diagonal 
Green functions in the Hamiltonians defined in Eq.~(\ref{eq:Chern}).
For $-4<m<0$, the system becomes a topological
insulator (Chern insulator) while 
it becomes a trivial band insulator for $m>0$.
At $m=0$, the quantum phase transition between 
the topological insulator and the trivial band insulator can occur ($m>0$). 
{Since the zeros traverse the bandgap,} 
the zeros of the Green functions {traverse} each other
in the topological insulator 
while they do not {traverse}
in the trivial band insulator.
(d) Brillouin zone and two-dimensional sphere.
In the topological insulator with the Chern number 1, 
there is a mapping from the Brillouin zone 
to the two-dimensional sphere, which covers the two-dimensional sphere at least once.
}
\label{fig:Chern}
\end{figure}%

\section{Two-dimensional Chern insulator (class A)}
\subsection{Zeros and poles in the Chern insulator}
To examine the behaviors of the zeros of the Green functions 
in topological insulators,
we employed a two-band model 
for a two-dimensional Chern insulator~\cite{Haldane_PRL1988} on a square lattice, defined as follows.
\begin{align}
&H_{\rm A}=\sum_{\nu=x,y}H_{\nu}+H_{\rm diag},\\
&H_{\nu}=\sum_{j}h_{\nu,j},\\
&{h_{\nu,j}=c_{j+\vec{e}_{\nu}}^{\dagger}T_{\nu}c_{j}+{\rm H.c.}},\\
&H_{\rm diag}=(2+m)\sum_{j}c_{j}^{\dagger}\sigma_{z}c_{j},
\label{eq:Chern}
\end{align}
where $c_{j}^{\dagger}$ ($c_{j}$)
represents the two-component fermion creation (annihilation)
operator that is defined on the site $j$ on the square lattice spanned by two orthogonal unit vectors $\vec{e}_{\nu=x,y}$.
The matrices $T_{\nu}$ are defined as
\begin{align}
T_{x}&=(-\sigma_{z}+{i}\sigma_{x})/2,\\
T_{y}&=(-\sigma_{z}+{i}\sigma_{y})/2,\\
\end{align}
where 
\begin{align}
\sigma_{x}=
\begin{pmatrix}
0 &~1 \\
1 &~0 \\
\end{pmatrix},
\sigma_{y}=
\begin{pmatrix}
0       &~-{i} \\
{i} &~0 \\
\end{pmatrix},
\sigma_{z}=
\begin{pmatrix}
1  &~0 \\
0  &~-1 \\
\end{pmatrix}.
\end{align}
It is noted that the system becomes a topological (Chern) insulator
for $-4<m<0$, and it becomes a trivial band insulator for $m>0$.

The Hamiltonian in the momentum space can be described as
\begin{align}
H_{\rm A}(\boldsymbol{k})&=
\begin{pmatrix} R_{z}(\boldsymbol{k}) & R_x(\boldsymbol{k})-iR_{y}(\boldsymbol{k}) \\ 
R_x(\boldsymbol{k})+iR_{y}(\boldsymbol{k})  & -R_{z}(\boldsymbol{k}) \end{pmatrix},\\
R_{x}(\boldsymbol{k})&=\sin{k_x},\\ 
R_{y}(\boldsymbol{k})&=\sin{k_y},\\
R_{z}(\boldsymbol{k})&=2+m-(\cos{k_x}+\cos{k_y}).
\end{align}
By diagonalizing the Hamiltonian, the eigenvalues (band dispersions) can be obtained as 
\begin{align}
E_{\pm}(\mb{k})&=\pm|R(\boldsymbol{k})| \notag \\
&=\sqrt{R_{x}(\boldsymbol{k})^2+R_{y}(\boldsymbol{k})^2+R_{z}(\boldsymbol{k})^2}.
\end{align}

Following the argument in Sec. II,
the zeros of the Green functions 
are represented by the eigenvalues of the 
minor matrix $M_{n}$.
For the 2$\times$2 matrices, 
the eigenvalues are  the diagonal components 
of the Hamiltonian.
Thus, the zeros of the $n$th {diagonal component of the} Green function ($\zeta_{n}$) are
represented by 
\begin{align}
\zeta_{0}(\boldsymbol{k}) &= M_{0}(\boldsymbol{k}) = -R_{z}(\boldsymbol{k}), \\
\zeta_{1}(\boldsymbol{k}) &= M_{1}(\boldsymbol{k}) = R_{z}(\boldsymbol{k}).
\end{align}

In Fig.~\ref{fig:Chern}(a)-(c), 
a plot is presented for the band dispersions 
and zeros of the Green functions for several values of $m$.
It was observed that the zeros of the Green function 
{traverse} each other in the topological insulator 
while they do not {traverse} in the trivial insulator. 
{The traverse is guaranteed by the traverse of the zeros.}
This feature of the zeros is in sharp contrast 
with the band dispersions (poles of Green functions). 
This is because it is impossible to distinguish the topological insulator and
band insulator based on only the band dispersions.
When considering the standard way to identify the topological insulators, 
it is necessary to examine the existence of the topological 
invariant or the gapless {edge} states.
However, as shown in Fig.~\ref{fig:Chern}, 
the {traverse} of the zeros of the Green functions in the bulk system
can be used to identify the topological insulator.
{We note that traverses of the zeros are robust
against the perturbations such as the next-nearest-neighbor transfers unless they do not
destroy the topological phases.}
The next subsection explains how the topological invariant can
guarantee the {traverse and the resultant} cross of the zeros of the Green functions.

\subsection{Relation with the Chern number}
This subsection shows that the
topological invariance 
(Chern number $\mathcal{C}$) guarantees the {traverse} of the zeros in the bandgap.
From {the eigenvector-eigenvalue identity} Eq.~(\ref{eq:EE}), an important relation can be obtained:
{a component of the eigenvector is zero ( $|\Psi_{j}^{(n)}(\vec{k})|^2=0$) $\leftrightarrow$ 
the zeros and poles of the Green function coincide ($\zeta(\vec{k}) = E(\vec{k})$)}.
These special points where $|\Psi_{j}^{(n)}|^2=0$ 
are called 
the vortex cores{~\cite{Kohmoto_1985ANN}}.
From the existence of such a special point, 
it is demonstrated how the topological invariant 
guarantees the {traverse} of the zeros.
If the Chern number $\mathcal{C}$ is nontrivial, for example, $\mathcal{C}=1$,
there is a one-to-one mapping 
from $(R_{x}(\vec{k}),R_{y}(\vec{k}),R_{z}(\vec{k}))$ to 
the two-dimensional sphere $S^{2}$,
which covers the two-dimensional sphere at least once.

Therefore, the points in the Brillouin zone exist
where $(R_{x}(\vec{k}),R_{y}(\vec{k}),R_{z}(\vec{k}))\propto(0,0,\pm 1)$.
{The zeros and poles coincide at these points.}
These points are located at $\Gamma=(0,0)$, $X=(\pi,0)$, $Y=(0,\pi)$, and $M=(\pi,\pi)$.
{For $R_{z}=1$ ($R_{z}=-1$), 
$\zeta_{0} = E_{1}$,\ 
and $\zeta_{1} = E_{0}$ ($\zeta_{0} = E_{0}$, and $\zeta_{1} = E_{1}$)
occur.} 
Thus, if $R_{z}$ takes $\pm$1 
(this is guaranteed by the existence of the topological invariant), 
$\zeta_{0}$ {traverses the band gap and crosses}
with $\zeta_{1}$ at least once.
Therefore, the existence of the 
Chern number induces the {traverse} of the
zeros in the topological insulator.

In contrast to the case of the topological insulator,
when the Chern number $\mathcal{C}$ is trivial ($\mathcal{C}=0$),
there exists $(R_{x}(\vec{k}),R_{y}(\vec{k}),R_{z}(\vec{k}))$ that does not
completely cover the two-dimensional sphere $S^{2}$. 
In other words, 
one such point is present
where $(R_{x},R_{y},R_{z})\propto(0,0,\pm 1)$ does not
exist in general. 
In general, 
the zeros of the Green function
do not traverse the bandgap and do not cross.
It is noted that the accidental crosses of the zeroes of the
Green function might occur even in trivial insulators.
An example is provided in Sec.~\ref{sec:CII}.

\subsection{Gauge invariance}
\label{sec:gauge}
As demonstrated, the {traverse} of the zeros
is guaranteed by the existence of the 
non-trivial topological invariant.
This result indicates that 
{the traverse of the zeros}
is gauge invariant. In this subsection, by explicitly 
performing the unitary transformation,
it is established that the 
{traverse} of the
zeros is gauge invariant.

By using the unitary matrix $U$,
a unitary transformation is performed as follows.
\begin{align}
\tilde{H}(\vec{k})&= U^{\dagger}H(\vec{k})U, \\
U&=
\begin{pmatrix}
u       &     ~~-v \\
v^{*}   &     ~~u 
\end{pmatrix}
\end{align}
where $u$ is a real number,
$v=v_{x}+iv_{y}$, and
$u^2+v_{x}^2+v_{y}^2=1$.
The explicit form of the transformed 
Hamiltonian is given as
\begin{align}
\tilde{H}(\vec{k})=
\begin{pmatrix}
aR_x+bR_y +cR_{z}  & h_{\rm off} \\
h_{\rm off}^{*}    & -(aR_x +bR_y +cR_{z}) 
\end{pmatrix},
\end{align}
where $a=2uv_{x}, b=-2uv_{y}, c = u^2-v_{x}^2-v_{y}^2$,
and $h_{\rm off}=(u^2-v^2)R_{x}-i(u^2+v^2)R_{y}+2u(-iv_{x}+v_y)R_{z}$.
It is noted that
$a^2+b^2+c^2=1$ is satisfied.
After the unitary transformation,
the zeros of the Green function are given as
\begin{align}
\zeta_{0}(\vec{k}) &= -(aR_x(\vec{k}) +bR_y(\vec{k}) +cR_z(\vec{k})) \\
\zeta_{1}(\vec{k}) &= (aR_x(\vec{k}) +bR_y(\vec{k}) +cR_z(\vec{k})).
\end{align}

For $R_{x}(\vec{k})=a|R|,R_{y}(\vec{k})=b|R|,R_{z}(\vec{k})=c|R|$,
the point where the zeros and poles coincide is given as
$\zeta_0 = E_1$ and $\zeta_1=E_0$.
For the antipodal point ($R_{x}=-a|R|, R_{y}=-b|R|, R_{z}=-c|R|$), 
the point is given by
$\zeta_1 = E_0$ and $\zeta_0=E_1$.
Because the existence of the topological invariant guarantees
that {$\vec{R}/|R|$} can cover the unit sphere,
$\vec{R}$ can take  $R_{x}=a|R|, R_{y}=b|R|, R_{z}=c|R|$ and 
the corresponding antipodal point.
Thus, the {traverse} of 
the zeros of the Green function
in Chern insulators
is gauge invariant.

\section{$\mathbb{Z}_2$ topological insulator in two and three dimensions (class AII)}
\label{sec:Z2}
In the previous section, as a canonical example of the
topological insulators, the zeros of the Green 
functions in the Chern insulators were analyzed.
{In this section, we examine} the zeros of the Green functions in
the $\mathbb{Z}_2$ (class AII) topological insulators in
two and three dimensions. 
It can be confirmed that the zeros {traverse} the bandgap in topological phases,
as demonstrated in the case of the Chern insulators.

\subsection{Kane-Mele model}
\begin{figure}[t!]
	\begin{center}
		\includegraphics[width=8cm]{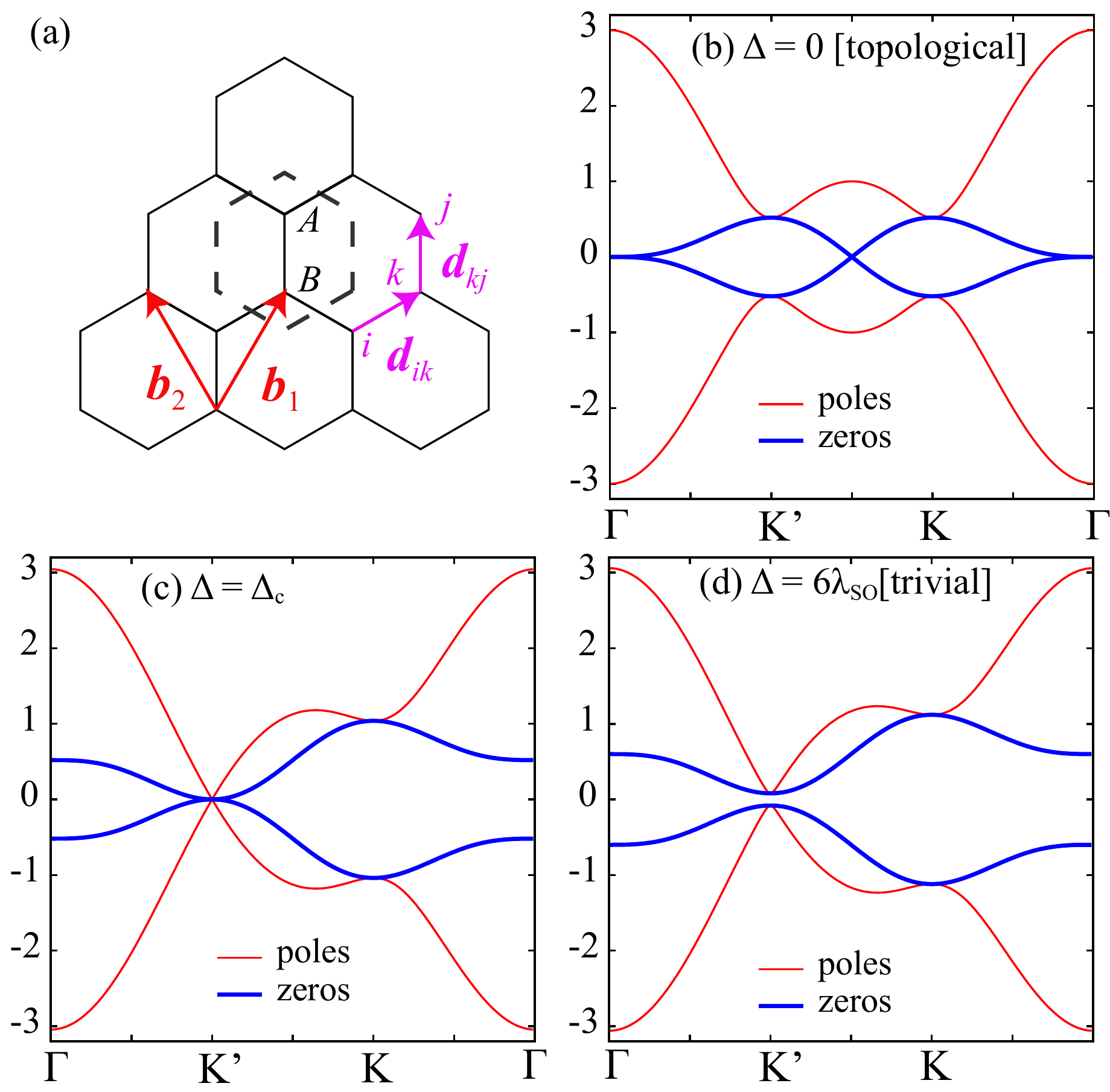}   
	\end{center}
\caption{(a)~Lattice structure and translational vectors used in the 
Kane-Mele model.~(b)–(d)~Band dispersions and the zeros of the Green functions
in the Kane-Mele model ($\lambda_{\rm SO}=0.1$).
The red lines display the band dispersions.
The thick blue lines represent the zeros of the Green function.
For simplicity, only the zeros and poles for the up spin are shown.
}
\label{fig:KM}
\end{figure}

As an example of the $\mathbb{Z}_2$ topological insulators,
the Kane-Mele model~\cite{KaneMele} was employed on the honeycomb lattice, which is defined as
\begin{align*}
H_{\rm KM} &= t\sum_{\la i,j\ra}\vc_{i}^{\dagger}\vc_{j}
+i\lambda_{\rm SO}\sum_{\la\la i,j \ra\ra}\nu_{ij}\vc_{i}^{\dagger}\sigma^{z}\vc_{j} \\ \notag
&+\sum_{i}\Delta_{i}\vc_{i}^{\dagger}\vc_{i}
+i\lambda_{R}\sum_{\la  i,j\ra}\vc_{i}^{\dagger}(\vec{\sigma}\times\frac{\vd_{ij}}{|\vd_{ij}|})_{z}\vc_{j}
\end{align*}
where $\vc_{i}^{\dagger}={(c_{i\uparrow}^{\dagger},c_{i\downarrow}^{\dagger})}$ and
$c_{i\sigma}^{\dagger}$ ($c_{i\sigma}$) is a creation (annihilation) 
operator of an electron with the spin $\sigma$ on site $i$.
Each parameter is defined as follows:
$t$ represents the nearest-neighbor hopping,
$\lambda_{\rm SO}$ represents the spin-orbit coupling, 
$\Delta_{i}$ is the staggered charge potential, and
$\lambda_{\rm R}$ represents the Rashba term.
$\vec{\sigma}$ is defined as $\vec{\sigma}=(\sigma^{x},\sigma^{y},\sigma^{z})^{t}$ where 
$\sigma^{x}$, $\sigma^{y}$, and $\sigma^{z}$ are Pauli matrices.
$\nu_{ij}$ is defined as $\nu_{ij}=(\vec{d}_{ik}\times\vec{d}_{kj})_z/|\vec{d}_{ik}\times\vec{d}_{kj}|$,
where $\vd_{ij}$ denotes the vector along
two bonds that the electron 
traverses from site $j$ to $i$ through $k$ (see Fig.~\ref{fig:KM}(a)).
For simplicity, we can first assume that the 
Rashba term is absent ($\lambda_{R}=0$).
The effects of the Rashba term are examined in the next subsection. 
For simplicity, $t=1$.

By performing a Fourier transformation, the
Hamiltonian without the Rashba terms in the momentum space 
can be described as
\begin{equation*}
H_{\sigma}(\boldsymbol{k})=\begin{pmatrix} R_{z,\sigma}(\boldsymbol{k}) & R_{x}(\boldsymbol{k})-iR_{y}(\boldsymbol{k}) \\ 
   R_{x}(\boldsymbol{k})+iR_{y}(\boldsymbol{k})& -R_{z,\sigma}(\boldsymbol{k}) \end{pmatrix},
\end{equation*}
where $R_{z,\sigma}(\boldsymbol{k})=\sigma E(\boldsymbol{k})+\Delta$ 
and $E(\mb{k})=4\lambda_{\rm SO}\sin(k_{x}/2)[\cos(k_x/2)-\cos(\sqrt{3}k_{y}/2)]$.
$R_{x}(\boldsymbol{k})$ [$R_{y}(\boldsymbol{k})$] is defined as
$R_{x}(\boldsymbol{k})={\rm Re}[E_{0}](\boldsymbol{k})$ 
[$R_{y}(\boldsymbol{k})={\rm Im}[E_{0}](\boldsymbol{k})$],
where $E_{0}(\boldsymbol{k})=1+e^{i\boldsymbol{k}\cdot\vb_{1}}+e^{i\boldsymbol{k}\cdot\vb_{2}}$ 
[$\vb_{1}=a(1/2,\sqrt{3}/2)$,$\vb_{2}=a(-1/2,\sqrt{3}/2)$]. For this investigation, the lattice constant was set as $a=1$.
By diagonalizing the Hamiltonian, the band dispersions can be obtained as 
\begin{align}
E_{0,\sigma}(\mb{k})&=\sqrt{R_{x}(\boldsymbol{k})^2+R_{y}(\boldsymbol{k})^2+R_{z,\sigma}(\boldsymbol{k})^2}, \notag \\
E_{1,\sigma}(\mb{k})&=-\sqrt{R_{x}(\boldsymbol{k})^2+R_{y}(\boldsymbol{k})^2+R_{z,\sigma}(\boldsymbol{k})^2}. 
\end{align}
The zeros of the Green functions are given as 
\begin{align}
\zeta_{0,\sigma}(\vec{k}) &= -R_{z,\sigma}(\vec{k}), \notag \\
\zeta_{1,\sigma}(\vec{k}) &= R_{z,\sigma}(\vec{k}).
\end{align}

In Fig.~\ref{fig:KM}(b)-(d), 
the band dispersions and zeros of the 
diagonal components of the Green functions are plotted for the up spin ($E_{i,\uparrow},\zeta_{i,\uparrow}$).
In the Kane-Mele model, 
spin-orbit coupling $\lambda_{\rm SO}$ induces the
$\mathbb{Z}_2$ topological insulators ~\cite{KaneMele}, while
the staggered charge potential $\Delta$ destroys the topological insulator.
By changing $\Delta$, the zeros of the Green function can be monitored in terms of how they
behave in the topological and trivial insulator. 

In Fig.~\ref{fig:KM}(b), it is demonstrated that the {zeros traverse the band gap}
in the $\mathbb{Z}_2$ topological insulator as in the
Chern insulator.
Because the $\mathbb{Z}_2$ topological insulator in the 
Kane-Mele model without the Rashba term is the
spin Chern insulator, the 
{traverses} of the zeros  
are guaranteed by the existence of the spin Chern number. 

By increasing the strength of the staggered charge potential $\Delta$, 
the insulator changes into the band insulator 
at $\Delta_{c}=3\sqrt{3}\lambda_{\rm SO}$.
At the transition point, the crosses of the zeros are lifted at
the Dirac point $K^{\prime}$ [Fig.~\ref{fig:KM}(c)] and 
there is no {traverse} in the trivial insulator [Fig.~\ref{fig:KM}(d)].
By using the same argument that is shown in Sec.~\ref{sec:gauge},
it is demonstrated the 
{traverse} of the zeros is gauge invariant.

\subsection{Effects of Rashba term}
This subsection considers the effects of the Rashba term.
By adding the Rashba term,
the Hamiltonian becomes $4\times4$ and
the spin Chern number is no longer well defined.
The Hamiltonian with the Rashba term is given as
\begin{align}
H_{\rm R}=\begin{pmatrix} 
R_{z,\uparrow}(\boldsymbol{k})               &  R_{\perp}(\boldsymbol{k})  & 0                   & z({\boldsymbol{k}}), \\ 
R_{\perp}(\boldsymbol{k})^{*} &   -R_{z,\uparrow}(\boldsymbol{k})              & w({\boldsymbol{k}})  & 0, \\
0                                            &   w({\boldsymbol{k}})^{*}         & R_{z,\downarrow}(\boldsymbol{k}) & R_{\perp}(\boldsymbol{k}),   \\
z({\boldsymbol{k}})^{*}                           &   0                          & R_{\perp}(\boldsymbol{k})^{*}    & -R_{z,\downarrow}(\boldsymbol{k})   \end{pmatrix},  \notag
\end{align}
where $R_{\perp}(\boldsymbol{k})=R_{x}(\vec{k})-iR_{y}(\vec{k})$, and
\begin{align*}
z({\boldsymbol{k}})&=\sqrt{3}\lambda_R[(-i(1-c_{x}c_{y})+c_{x}s_{y})/\sqrt{3}+s_{x}s_{y}+is_{x}c_{y}], \\
w({\boldsymbol{k}})&=\sqrt{3}\lambda_R[(i(1-c_{x}c_{y})+c_{x}s_{y})/\sqrt{3}-s_{x}s_{y}+is_{x}c_{y}].
\end{align*}
Here, we define $c_{x}$ [$s_{x}$] as $\cos(k_{x}/2)$ [$\sin(k_{x}/2)$]
and  $c_{y}$ [$s_{y}$] as $\cos(\sqrt{3}k_{y}/2)$ [$\sin(\sqrt{3}k_{y}/2)$].
For the Hamiltonian, because it is difficult to 
obtain the simple analytical form of the eigenvalues and the zeros,
they were numerically obtained.

\begin{figure}[t!]
	\begin{center}
		\includegraphics[width=8cm]{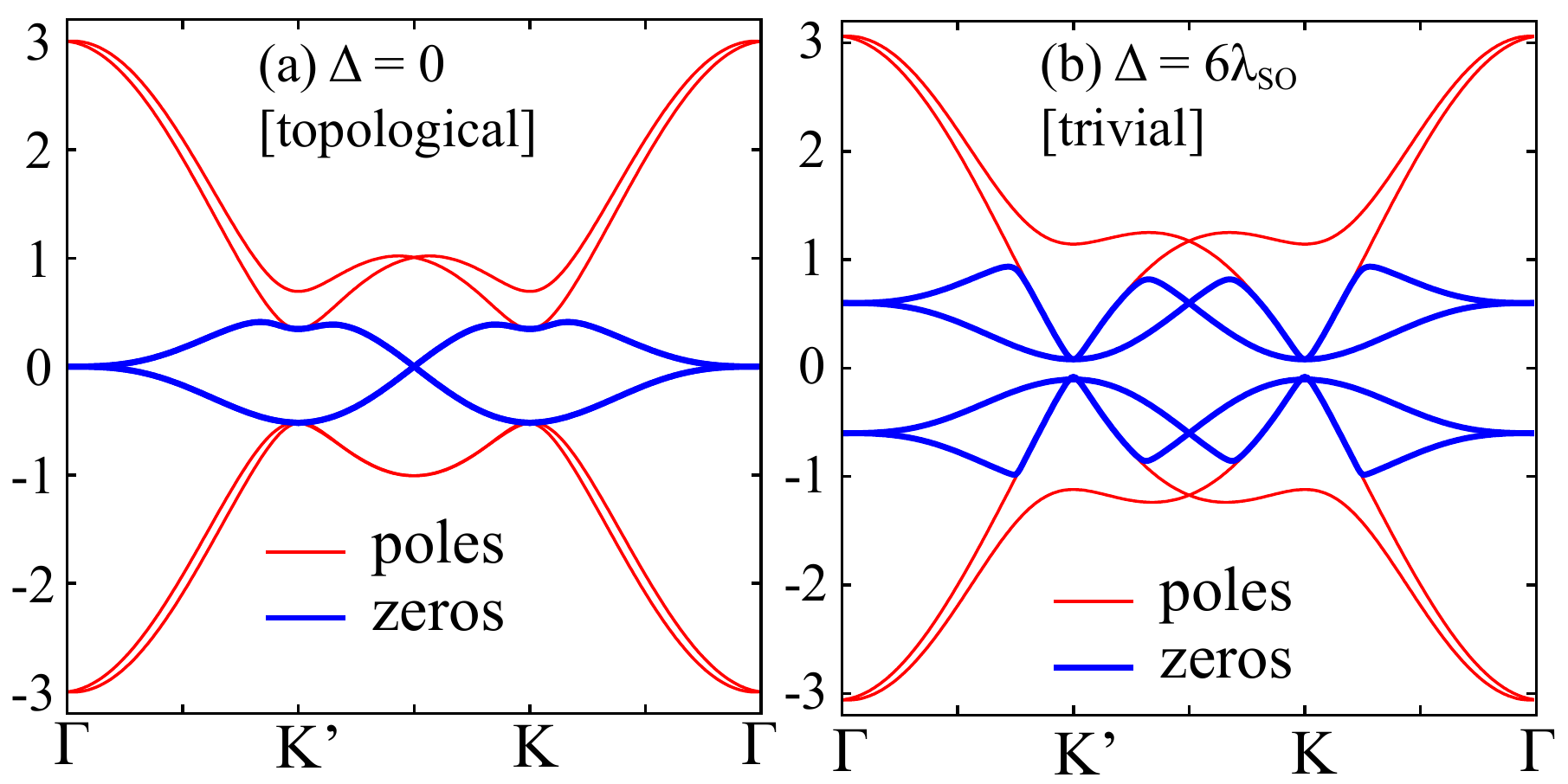}   
	\end{center}
\caption{~Band dispersions and zeros of the Green functions
in the Kane-Mele model with the Rashba term ($\lambda_{\rm SO}=0.1$ and $\lambda_{\rm R}=0.1/\sqrt{3}$)
for (a) the topological insulator and (b) the trivial insulator.
The red lines display the band dispersions.
The thick blue lines represent the zeros of the Green function.
Only the zeros of Green functions in the bandgap are presented.
}
\label{fig:Rashba}
\end{figure}%

Even under the existence of the Rashba term, 
the 
{traverses} of zeros appear in the topological insulator
as shown in Fig.~\ref{fig:Rashba}(a) while
the 
{traverses} disappear 
in the trivial insulators occur as shown in Fig.~\ref{fig:Rashba}(b).
At the $K^{\prime}$ point, 
the first and fourth zeros in the bandgap 
coincide with the valence bands, 
while the second and third zeros 
coincide with the conduction bands. 
At the $K$ point, since the sign of the mass term is 
the opposite from the $K^{\prime}$ point, 
the opposite occurs.
To connect each zero consistently,
the first and the fourth zeros 
and the second and third zeros should traverse
the bandgap and should cross at least 
once as shown in Fig.~\ref{fig:Rashba}(a).
These results indicate that the zeros of the Green function are also
useful for characterizing the $\mathbb{Z}_2$ 
topological insulator.

\subsection{$\mathbb{Z}_2$ topological insulator in three dimensions (class AII)}
This subsection considers the zeros of the Green functions
in the three-dimensional topological insulators.
A typical model of the three-dimensional topological insulator on the cubic lattice~\cite{Qi_RRB2008,Liu_PRB2010}
is given as follows.
\begin{align}
R_{0} &= M_{0}+3-(\cos{k_x}+\cos{k_y}+\cos{k_z}),                       \\
R_{1}  &= \sin{k_x},  \\
R_{2}  &= \sin{k_y},  \\
R_{3}  &= \sin{k_z},  \\
H     &= R_0\alpha^{0}+R_{1}\alpha^{1}+R_{2}\alpha^{2}+R_{3}\alpha^{3} \\
      &=
\begin{pmatrix}
R_{0}         & 0             & R_{3}         & R_{1}-iR_{2}  \\
0             & R_{0}         & R_{1}+iR_{2}  & -R_{3}  \\
R_{3}         & R_{1}-iR_{2}  & -R_{0}        & 0  \\
R_{1}+iR_{2}  & -R_{3}        & 0             & -R_{0}  \\
\end{pmatrix},
\label{eq:3DTI}
\end{align}
{where $\alpha^{0}=1\otimes\sigma_{z}$, 
$\alpha^{1}=\sigma_{x}\otimes\sigma_{x}$,
$\alpha^{2}=\sigma_{y}\otimes\sigma_{x}$, and
$\alpha^{3}=\sigma_{z}\otimes\sigma_{x}$.}
In this model, 
the $\mathbb{Z}_2$ topological insulator is achieved
for $-2<M_{0}<0$ and $-6 < M_{0}<-4$ 
while the trivial insulator is realized
for $M_{0}<-6$ and $M_{0}>0$.  
The weak $\mathbb{Z}_2$ topological insulator is obtained
for $-4<M_{0}<-2$. 
{We note that this model has additional chiral symmetry 
because $\alpha^{4}=\alpha^{0}\alpha^{1}\alpha^{2}\alpha^{3}=1\otimes\sigma_{y}$ 
anticommutes with $H$~\cite{Qi_RRB2008}. }

The eigenvalues are given as follows.
\begin{align}
E_{0}&=\sqrt{R_{0}^2+R_{1}^2+R_{2}^2+R_{3}^2}, \\
E_{1}&=-\sqrt{R_{0}^2+R_{1}^2+R_{2}^2+R_{3}^2}
\end{align}
It can be noted that the eigenvalues are doubly degenerate.
The zeros of the Green function in the bandgap are 
given as follows.
\begin{align}
\zeta_{0}&=-R_{0}, \notag \\
\zeta_{1}&=R_{0}.
\end{align}

From these expressions, it can be shown that
the zeros and the poles coincide 
at eight time-reversal points, such as the $\Gamma$, X, and Z points.
To analyze the 
{traverses} of the zeros, the following was defined.
\begin{align}
\eta(\vec{k}) = {\rm sign}(\zeta_{1}(\vec{k})).
\end{align}
If $\eta(\vec{k})=1$ ($\eta(\vec{k})=-1$) at the time reversal point, 
$\zeta_{1}(\vec{k})$ coincides with $E_{0}(\vec{k})$ ($E_{1}(\vec{k})$).
$\eta(\vec{k})$ at the time-reversal points are shown in the inset.

\begin{figure}[t!]
	\begin{center}
		\includegraphics[width=8cm]{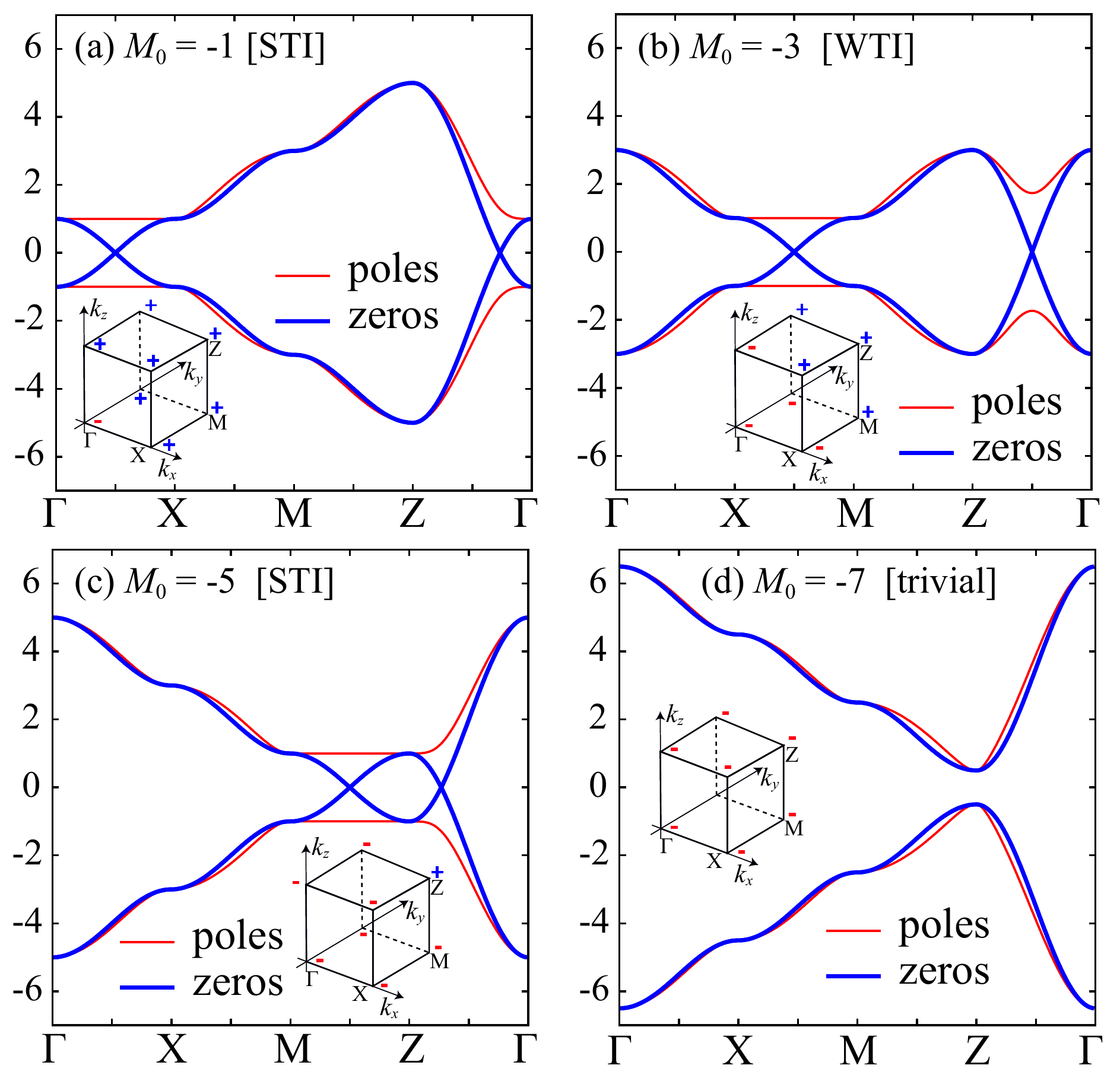}   
	\end{center}
\caption{~Band dispersions and zeros of the Green functions
in the Hamiltonian defined in Eq.~(\ref{eq:3DTI}) for
(a) a strong topological insulator,
~(b) a weak topological insulator,
~(c) a strong topological insulator,
~(d) and a trivial insulator.
The red lines display the band dispersions, and
the thick blue lines represent the zeros of the Green function.
In the inset, 
$\eta(\vec{k})$ is shown for the time-reversal points.
STI (WTI) is an abbreviation for a strong (weak) topological insulator.
}
\label{fig:3DTI}
\end{figure}

Because $\zeta_{0}(\vec{k})=-\zeta_{1}(\vec{k})$,
they 
{traverse} between $\eta(\vec{k})=1$ and $\eta(\vec{k})=-1$. 
For example, for $M_{0}=-1$ [Fig.~\ref{fig:3DTI}(a)], 
they 
{traverse} between $\Gamma$ point and X point. 
Even for the weak topological region, 
they 
{traverse} as shown in Fig.~\ref{fig:3DTI}(a) and
they do not {traverse} in the trivial insulators.
These results indicate that the {traverses} of the
zeros can be useful for detecting the topological phases
for strong and weak topological insulators since 
the {traverses} are guaranteed by the existence of the 
{band inversions, as we detail in the next section.}

\section{Relation with band inversions}
{In this section, we explain why the {traverses} of 
the zeros universally appear in the topological insulators.
In the topological insulators, the band inversions 
generally occur.
We denote that the occupied (unoccupied) eigenstates with the index $i\leq0$ ($i>0$). 
The eigenvectors and eigenvalues are given by
\begin{align}
H(\vec{k})\vec{e}_{i}(\vec{k}) = E_{i}(\vec{k})\vec{e}_{i}(\vec{k}).
\end{align}
Here, $E_{0}(\vec{k})$ ($E_{1}(\vec{k})$)
is the highest occupied (lowest unoccupied) eigenvalues.
We assume that the band gap exists between $E_{0}$ and $E_{1}$, i.e.,
$E_{0}(\vec{k})<E_{1}(\vec{k})$ {at any $\vec{k}$}.}
Here, we define the band inversion as follows:
A single pair of momenta $\vec{k}$, $(\vec{k}_{0},\vec{k}_{1})$ 
exists such that
one occupied eigenvector at $\vec{k}_{1}$ [$\vec{e}_{n_{0}}(\vec{k}_{1})$]
is orthogonal to all the occupied eigenvectors at $\vec{k}_{0}$, i.e.,
\begin{align}
\vec{e}_{i}(\vec{k}_{0})^{\dagger}\vec{e}_{n_{0}}(\vec{k}_{1}) = 0~~i\in[-(N-1),0],
\label{eq:bandinversion}
\end{align}
where $N$ is the number of the occupied bands.
We detail the relation with the band inversion and
the existence of the Chern and $Z_{2}$ topological number
in Appendix~\ref{ap:bandinversion}.

\begin{figure}[t!]
	\begin{center}
	\includegraphics[width=6cm,clip]{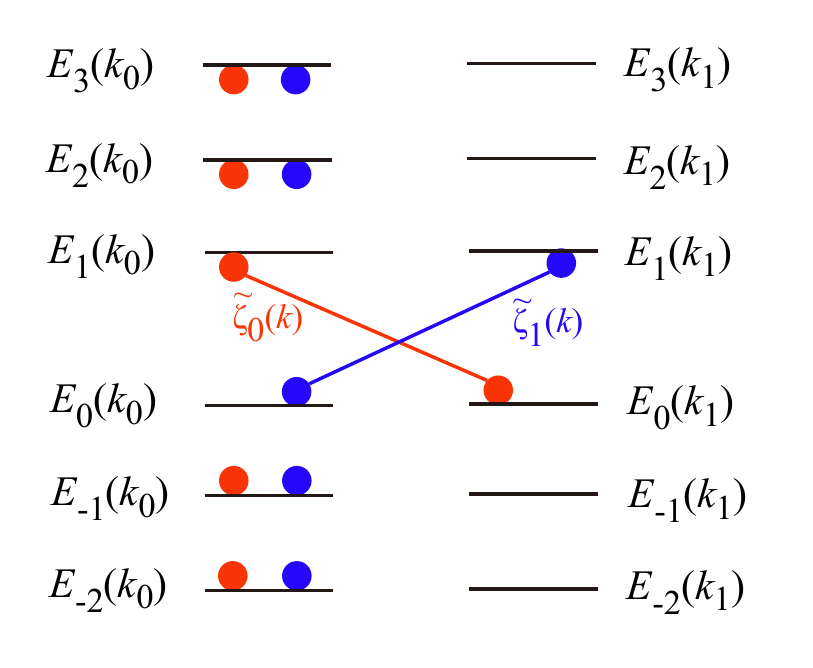}   
	\end{center}
\caption{Schematic picture of the eigenvalues and
the zeros of the Green function around the band gap under
the unitary transformation defined in Eq.~(\ref{eq:unitary}).
The circles represent the points where the zeros and poles coincide.
At $\vec{k}_{0}$, all the zeros coincide with the poles
as shown in the left panel.
If band inversions occur at $\vec{k}_{1}$, 
the zeros traverse the band gap as shown in the right panel.
We note that the zeros outside the band gap do not necessarily
coincide with the poles at $\vec{k}_{1}$. Thus, we do not 
show the position of the zeros outside the band gap.
}
\label{fig:Schem}
\end{figure}

{Then, using the eigenvectors at a certain $\vec{k}$ point ($\vec{k}=\vec{k}_0$), 
we construct the unitary matrix
\begin{align}
U^{\dagger}=
\begin{pmatrix}
\vec{e}_{0}(\vec{k}_{0})^{\dagger}\\
\vec{e}_{1}(\vec{k}_{0})^{\dagger}\\
\vec{e}_{-1}(\vec{k}_{0})^{\dagger}\\
\vec{e}_{2}(\vec{k}_{0})^{\dagger}\\
\vec{e}_{-2}(\vec{k}_{0})^{\dagger}\\
\vdots\\
\end{pmatrix},
\label{eq:unitary}
\end{align}
We note that the eigenvectors of 
the transformed Hamiltonian $\tilde{H}(\vec{k})=U^{\dagger}H(\vec{k})U$
are given by $\tilde{\vec{e}}_{i}(\vec{k})=U^{\dagger}\vec{e}_{i}(\vec{k})$.
For example, the lowest two eigenvectors are transformed as
\begin{align}
&\tilde{\vec{e}}_{0}(\vec{k}_0)=U^{\dagger}\vec{e}_{0}(\vec{k}_{0})=(1,0,0,\dots,0)^{t},\\
&\tilde{\vec{e}}_{1}(\vec{k}_0)=U^{\dagger}\vec{e}_{1}(\vec{k}_{0})=(0,1,0,\dots,0)^{t}.
\end{align}
Thus, under this unitary transformation, the zeros of the 0th (1st) 
component of the diagonal Green functions $\tilde{\zeta}_{0}$ ($\tilde{\zeta}_{1}$) 
coincide with the lowest unoccupied 
(highest occupied band) due to the
eigenvector-eigenvalue identity 
{Eq.~(\ref{eq:EE})}.
This situation is schematically shown in Fig.~\ref{fig:Schem}.}

{Then, we will show that the zeros of the Green function,
\begin{align}
{\tilde{G}=[\omega I-\tilde{H}(\vec{k})]^{-1},}
\end{align}
traverse the band gap in the topological phase, where the band inversion occurs.}

{{Here, we assume
the band inversion between}
$\vec{k}_{0}$ and $\vec{k}_{1}$, which is defined in
Eq.~(\ref{eq:bandinversion}).
Without loss of generality, we can take $n_{0}=0$.
{Then}, we obtain
\begin{align}
&\tilde{\vec{e}}_{0}(\vec{k}_1)=U^{\dagger}\vec{e}_{0}(\vec{k}_{1})=(0,c,0,\dots)^{t},
\end{align}
where $c$ is a non-zero number.
This {illustrates}
that $\tilde{\zeta}_0$ traverses the band gap, i.e.,
$\tilde{\zeta}_{0}$ coincides with $E_{1}(\vec{k}_0)$ 
at $\vec{k}_0$ and
coincides with $E_{0}(\vec{k}_1)$ at $\vec{k}_1$ (see Fig.~\ref{fig:Schem}). 
We can also show that $\tilde{\zeta}_{1}$ also traverses the band gap
since the band inversion also occurs in the unoccupied states 
according to the argument given in Appendix~\ref{ap:bandinversion}.
This result shows the band inversion
{induces}
the traverses of the zeros in the band gap.
In Appendix.~\ref{ap:traverse},
{a detailed condition
for the traverse of the zeros is given}.
According to Eq.~(\ref{eq:inversion}),
when $\vec{e}_{0}(\vec{k}_{1})$ is orthogonal to $\vec{e}_{1}(\vec{k}_{0})$ and
mainly consists of the unoccupied eigenvectors 
at $\vec{k}_{0}$, $\tilde{\zeta}_0$ traverses the band gap.
We note that this condition can be 
regarded as a generalization of the band inversion defined in Eq.~(\ref{eq:bandinversion})}.

This argument demonstrates that 
the traverse and the resultant crosses 
of the zeros always appear in the topological phase
by taking the unitary transformation defined in Eq.~(\ref{eq:unitary}).
As we will show in the next section, using the unitary transformation,
the {traverses} {of} zeros appear in the topological insulators with chiral symmetry.
Before that, we explain why the {traverses} of the zeros appear 
in the Chern 
insulator {\it without} taking the 
unitary transformation.

For the Chern insulator, the Chern number is defined as~\cite{Kohmoto_1985ANN}
\begin{align}
&\mathcal{C}=\sum_{n=0}^{N-1}\frac{1}{2\pi}\int_{\rm BZ}d\vec{k}[\nabla\times\vec{a}_{-n}(\vec{k})], \\
&a_{-n}(\vec{k}) = -i\vec{e}_{-n}(\vec{k})^{\dagger}\frac{\partial\vec{e}_{-n}(\vec{k})}{\partial \vec{k}}.
\end{align}
It is known if the Berry connection 
$a_{-n}(\vec{k})$ is smooth over the Brillouin zone,
the Chern number becomes zero according to the Stokes theorem~\cite{Kohmoto_1985ANN}.
Therefore, if the Chern number is non-zero, the point where the phase
of the eigenvector is not well defined should exist (the point is called vortex core).
At that point, one component of the eigenvector becomes 0, i.e.,
the zeros of the Green function ($\zeta_{i}$)
coincides with the eigenvalue $E_{0}$.
Due to the band inversion, the same coincidence should occur for
the unoccupied band, i.e., 
$\zeta_{i}$ coincides with  $E_{1}$ at the different momentum.
Therefore the zeros of the Green functions ($\zeta_{i}$) traverse the band gap.
Since another zero ($\zeta_{j}$) also traverses the band gap, 
the zeros cross each other.
This shows that the existence of the vortex core in any gauges is 
the reason why the {traverses} of the zeros appear in the Chern insulator
without taking the unitary transformation.
We note that the similar discussion can be applied to
the four-dimensional $2\mathbb{Z}$ topological insulators (class AI)
because the topological invariant is characterized by the second Chern number.
In fact, as shown in Appendix~\ref{ap:AI}, the traverses of the
zeros appear in AI topological insulator without the unitary transformation.

\section{Topological insulators with chiral symmetry (class AIII,BDI,CII)}
\subsection{Choice of proper gauge}
Thus far, we examined the zeros of the Green functions
in the topological insulators {\it without} chiral symmetry.
Here, we examine the behavior of the zeros of the Green functions
in the topological insulators {\it with} chiral symmetry.

For the Hamiltonian that has chiral symmetry,
the unitary matrix $\Gamma$ that anti-commutes with the 
Hamiltonian always exists, i.e.,
\begin{align}
H\Gamma+\Gamma H=0.
\end{align}
From this, the Hamiltonian with chiral symmetry and 
its Green function at zero energy
can be described as a block off-diagonal form as demonstrated below.
\begin{align}
&H=
\begin{pmatrix}
0           & ~~Q \\
Q^{\dagger} & ~~0
\end{pmatrix}, \label{eq:ham_chiral}\\
&G(\omega=0)=-H^{-1}=
\begin{pmatrix}
0       & ~~-(Q^{\dagger})^{-1} \\
-Q^{-1} & ~~0
\end{pmatrix}.
\label{eq:green_chiral}
\end{align}
In this representation,
the unitary matrix $\Gamma$ has a diagonal form, which is given as
\begin{align}
\Gamma=
\begin{pmatrix}
I           & ~~0 \\
0           & ~~-I
\end{pmatrix}.
\end{align}

From Eq.~(\ref{eq:green_chiral}),
it can be observed that the zeros of the bulk Green function in the bandgap
always exist at $\omega=0$ for the 
topological and trivial insulators.
This result indicates that it is impossible to
distinguish the topological phase from 
the trivial phases if we take the block off-diagonal form of
the Hamiltonians.
{We also note that the vortex core is apparently absent
in this gauge since the zero can not coincide with the poles
if the finite gap exists. In other words, a component of 
the eigenvectors are always non-zero and the smooth gauge can be taken.
This feature is in sharp contrast with the topological insulators without chiral symmetry
such as the Chern insulators.
Thus, to identify the topological phase with chiral symmetry,
it is necessary to perform the proper unitary transformation.
By taking the unitary transformation defined in Eq.~(\ref{eq:unitary}),
as we show later, the traverses of the zeros appear in the
topological insulators with chiral symmetry because 
the band inversions occur in the topological phases.
}

\subsection{BDI and AIII topological insulators in one-dimension}
As an example of the BDI and AIII topological insulators,
the one-dimensional {Hamiltonian~\cite{Su_PRL1979,Velasco_PRL_2017}} can be considered, which is given as
\begin{align}
H=
\begin{pmatrix}
0         & R_x-i R_y \\
R_x+i R_y & 0
\end{pmatrix},
\label{eq:chiral}
\end{align}
where $R_x=1+\gamma\cos{(x-\delta)}$ and $R_{y}=\gamma\sin{(x-\delta)}$.
When $\gamma>1$ and $\delta=0$, 
this system becomes a topological insulator ($d=1$, BDI).
If $\delta\neq0$ and $\gamma>1$,
the system becomes an AIII topological 
insulator~\cite{Velasco_PRL_2017}.

{
The eigenvalues and eigenvectors of the Hamiltonian are given as follows.
\begin{align}
&E_{\pm}=\pm\sqrt{R_{x}^2+R_{y}^2},\\
&\vec{e}_{\pm}=\frac{1}{\sqrt{2}}
\begin{pmatrix}
\frac{\alpha^{*}}{|\alpha|} \\
\pm1 \\
\end{pmatrix},
\end{align}
where $\alpha=R_{x}+iR_{y}$.
Following the discussion above, by taking $k_0=\delta$ so as to
satisfy the relation $R_{y}(k_0)=\sin(k_{0}-\delta)=0$,
the unitary transformation is given by
\begin{align}
U^{\dagger}=
\begin{pmatrix}
\vec{e}_{-}(k_0)^{\dagger} \\
\vec{e}_{+}(k_0)^{\dagger} 
\end{pmatrix}
=\frac{1}{\sqrt{2}}
\begin{pmatrix}
1  &~~-1 \\
1  &~~1 \\
\end{pmatrix}.
\end{align}
Using the unitary matrix,
the transformed Hamiltonian
is obtained as follows.
\begin{align}
\tilde{H}=U^{\dagger}HU=
\begin{pmatrix}
R_x           & ~~iR_{y}\\
-iR_{y}      & ~~-R_x
\end{pmatrix}.
\end{align}
The zeros of the Green function for the transformed Hamiltonian 
$\tilde{H}$ are 
given as follows.
\begin{align}
\tilde{\zeta}_{0} &= -R_{x}, \notag \\
\tilde{\zeta}_{1} &=  R_{x}.
\end{align}
}

\begin{figure}[t!]
	\begin{center}
		\includegraphics[width=8cm]{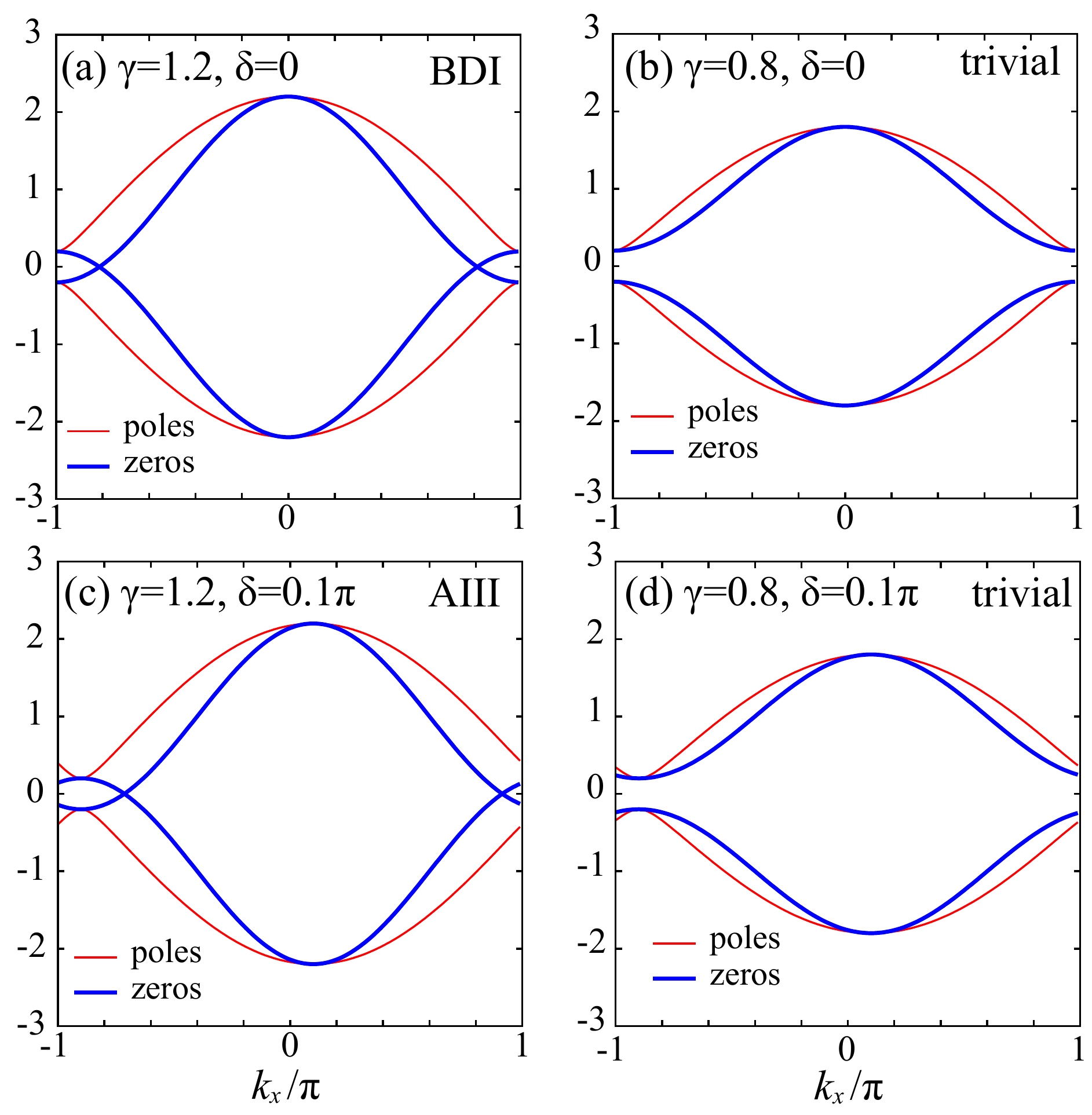}   
	\end{center}
\caption{~Band dispersions and zeros of the Green functions
in the model Hamiltonian for (a),(b) BDI, and (c),(d) AII topological insulators,
which is defined in Eq.~(\ref{eq:chiral}).
The red lines display the band dispersions, and
the thick blue lines represent the zeros of the Green function.
The zeros traverse the bandgap for the BDI 
and AIII topological insulators.
}
\label{fig:chiral}
\end{figure}

For the BDI topological insulator ($\delta=0,\gamma>1$),
the zeros {traverse} the bandgap as shown in Fig.~\ref{fig:chiral}(a)
whereas the zeros do not {traverse} in the trivial insulator.
It can be confirmed that the same behavior occurs in the
AIII topological insulators 
as shown in Fig.~\ref{fig:chiral}(c).

\subsection{CII topological insulator in one-dimension}
\label{sec:CII}
As an example of the CII topological insulators,
the following four-orbital models were applied~\cite{Liu_IOP_2019}.
\begin{align}
H=
\begin{pmatrix}
0         &~~0       &~~R_x    &~~iA+R_y \\
0         &~~0       &~~iA+R_y &~~R_x    \\
R_{x}     &~~-iA+R_y &~~0      &~~0      \\
-iA+R_y   &~~R_{x}   &~~0      &~~0 
\end{pmatrix},
\end{align}
where $R_x=t_{\perp}$, $R_{y}=2t\sin{k_x}$, and
$A=s+2s^{\prime}\cos{k_x}$.
When $(t_{\perp}/t)^2+(s/s^{\prime})^2<4$, 
this system becomes a topological insulator ($d=1$, CII).
The eigenvalues of the Hamiltonian are given as follows.
\begin{align}
E_{0}&=-\sqrt{(R_{x}+R_{y})^2+A^2}, \\
E_{1}&=-\sqrt{(R_{x}-R_{y})^2+A^2}, \\
E_{2}&=\sqrt{(R_{x}-R_{y})^2+A^2}, \\
E_{3}&=\sqrt{(R_{x}+R_{y})^2+A^2}.
\end{align}

\begin{figure}[t!]
	\begin{center}
		\includegraphics[width=8cm]{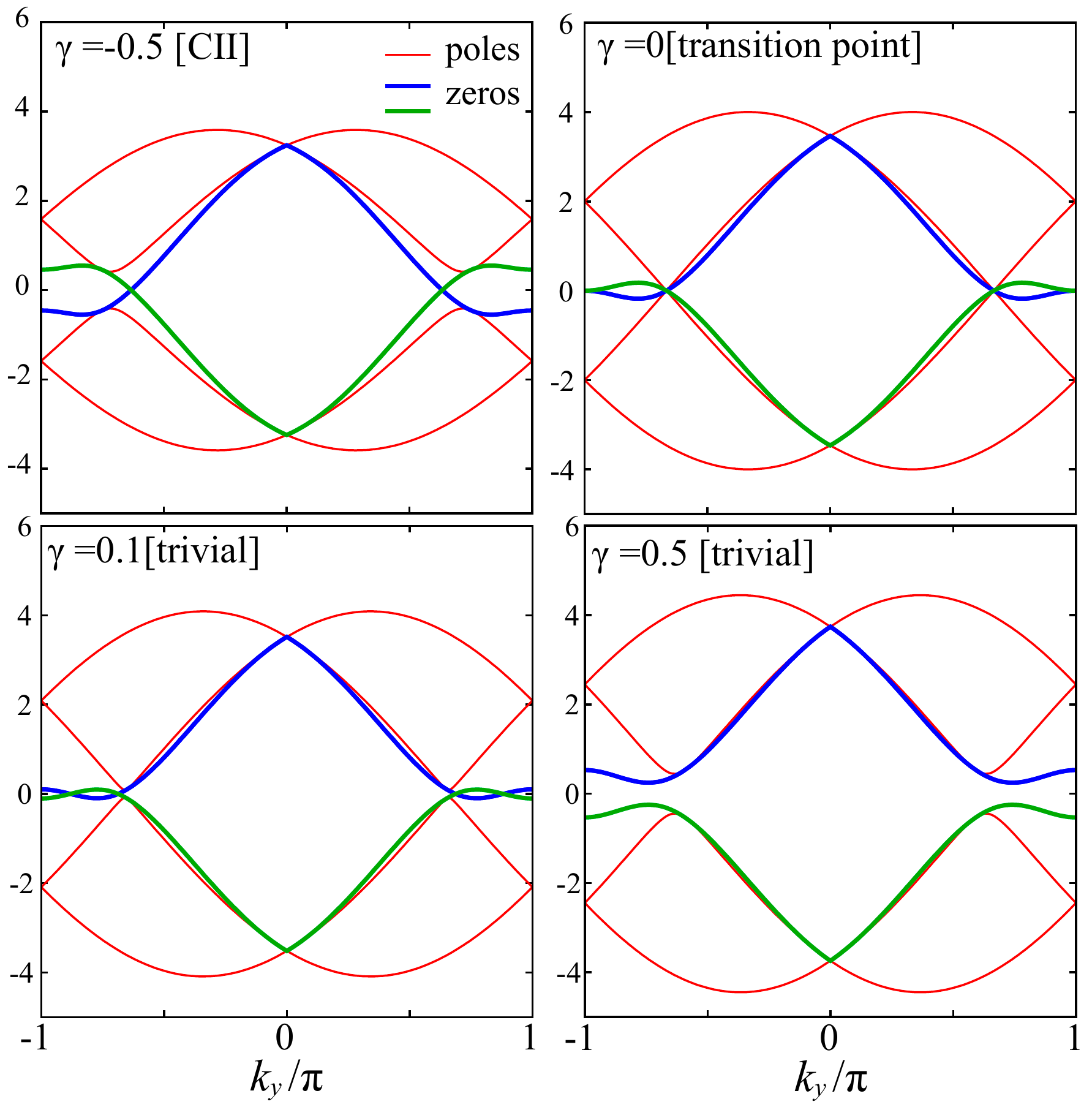}   
	\end{center}
\caption{~Band dispersions and zeros of the Green functions
in the model {of the CII topological insulators} 
for the
(a) topological insulator,
~(b) transition point,
~(c) trivial insulator {near the phase transition},
~(d) and the trivial insulator.
For this investigation, $t=1, s=s^{\prime}=1$, and 
$t_{\perp}=\sqrt{4-(s/s^{\prime})^2}+\gamma$.
For $\gamma<0$ ($\gamma>0$), 
the system becomes a topological (trivial) insulator.
The red lines display the band dispersions, and the
thick blue lines represent the zeros of the Green function.
}
\label{fig:CII}
\end{figure}

{
Following the discussion above, by tanking $k_{0}=0$,
we obtain the unitary matrix 
by numerically diagonalizing the Hamiltonian.
Using the unitary matrix, 
we numerically obtain the zeros of the Green function of 
$\tilde{H}=U^{\dagger}HU$.
The zeros are shown in Fig.~\ref{fig:CII}.
We note that the eigenstates are degenerate at $k=0$ and 
the form of the unitary transformation is not unique.
Nonetheless, the traverses of the zero are guaranteed by 
the existence of the band inversion as we showed above.
}

As illustrated in Fig.~\ref{fig:CII}(a),
in the CII topological insulator,
the zeros of the Green functions 
{traverse} the bandgap.
Basically, the crosses vanish in the trivial insulator 
sufficiently away from the transition point.
However, the accidental crosses of the zeros 
still survive, even in the trivial 
insulators as shown in Fig.~\ref{fig:CII}(c)
near the transition point.
{
These crosses are not guaranteed by the traverse of the zeros
and we can remove the accidental crosses without gap closing.
}

\section{Higher-order topological insulators}
As demonstrated in the previous sections,
the zeros {traverse} in the topological insulators
because the gapless edge states inevitably induce the 
{traverse} of the zeros.
Recently, new classes of the topological insulators have been found, 
such as the higher-order topological 
insulators~\cite{Benalcazar_Science2017,Schindler_SA2018}.
In the higher-order topological insulators in $n$ dimensions,
the gapless edge states appear below $n-2$ dimensions.
For example,
the three-dimensional higher-order topological insulators 
have one-dimensional edge states and
the two-dimensional  higher-order topological insulators 
have zero-dimensional edge states.
{In this section, we demonstrate} that the {traverses} of the zeros in the bulk system
also appear in the higher-order topological insulators
{because the band inversion occurs.}

\subsection{Higher-order topological insulators in three dimensions}
A model of the three-dimensional 
higher-order topological insulator on the cubic lattice
is described as follows~\cite{Schindler_SA2018}.
\begin{align}
R_{0} &= M+(\cos{k_x}+\cos{k_y}+\cos{k_z}),                       \\
R_{1}  &= \sin{k_x},  \\
R_{2}  &= \sin{k_y},  \\
R_{3}  &= \sin{k_z},  \\
R_{4}  &= \Delta_{2}(\cos{k_x}-\cos{k_y}),  \\
H     &= R_0\alpha^{0}+R_{1}\alpha^{1}+R_{2}\alpha^{2}+R_{3}\alpha^{3}+R_{4}\alpha^{4} \\
      &=
\begin{pmatrix}
R_{0}         & ~0             & R_{3}-iR_{4}  & ~R_{1}-iR_{2}  \\
0             & ~R_{0}         & R_{1}+iR_{2}  & ~-R_{3}-iR_{4}   \\
R_{3}+iR_{4}  & ~R_{1}-iR_{2}  & -R_{0}        & ~0  \\
R_{1}+iR_{2}  & ~-R_{3}+iR_{4} & 0             & ~-R_{0}  \\
\end{pmatrix}
\label{eq:3DHOT}
\end{align}
The eigenvalues are given as
\begin{align}
E_{0}&= \sqrt{R_{0}^2+R_{1}^2+R_{2}^2+R_{3}^2+R_{4}^2}, \\
E_{1}&=-\sqrt{R_{0}^2+R_{1}^2+R_{2}^2+R_{3}^2+R_{4}^2},
\end{align}
and
the zeros are given as
\begin{align}
\zeta_{0}&=-R_{0}, \\
\zeta_{1}&= R_{0}.
\end{align}
It can be noted that without $R_{4}$, 
this model is the same as that of the three-dimensional
$\mathbb{Z}_2$ topological insulators defined in Eq.~(\ref{eq:3DTI}).
It is shown that the system changes from the $\mathbb{Z}_2$ topological insulator into a higher-order 
topological insulator when 
$R_{4}$ is added ~\cite{Schindler_SA2018}.

\begin{figure}[t!]
	\begin{center}
		\includegraphics[width=8cm]{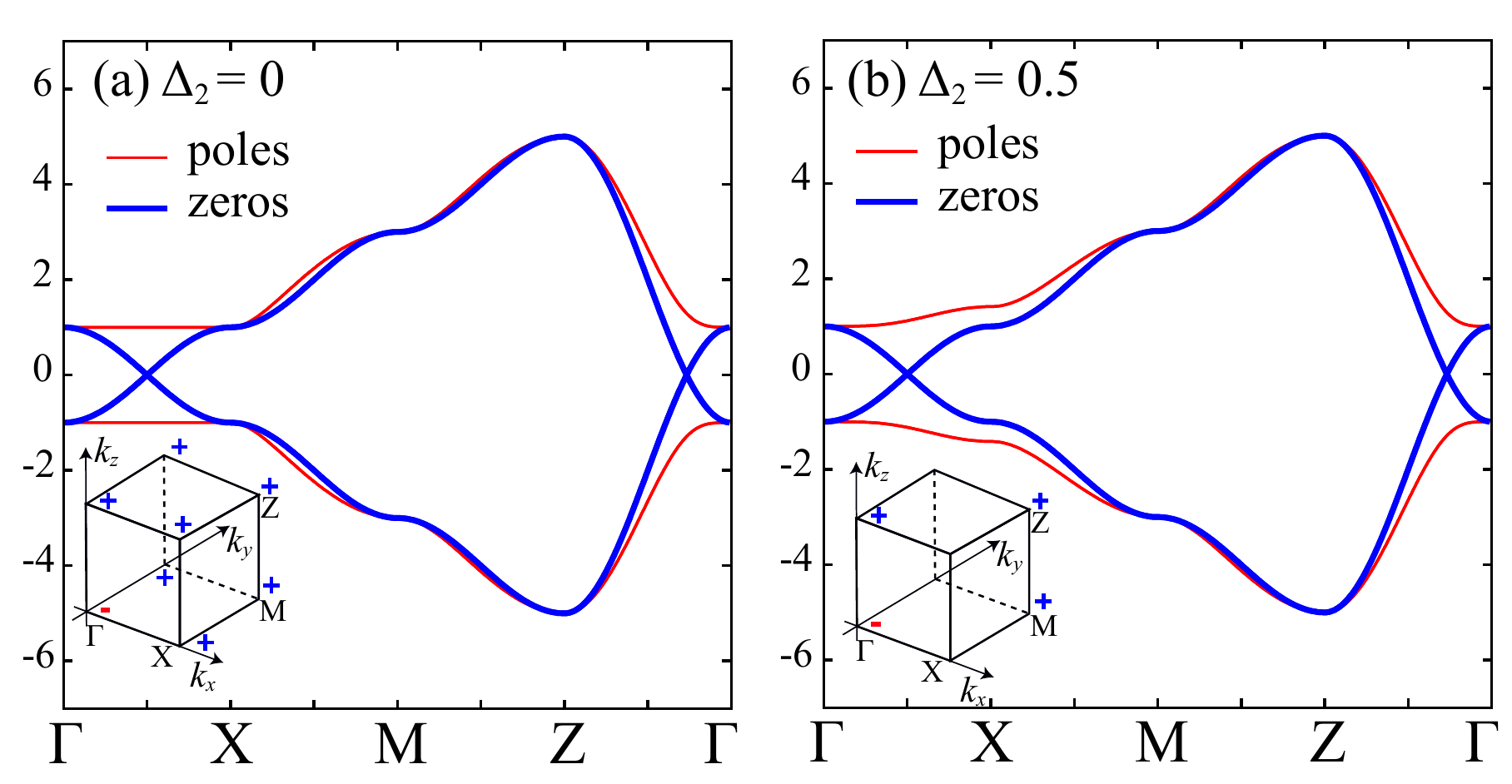}   
	\end{center}
\caption{~Band dispersions (thin red curves) 
and zeros (thick blue curves) 
of the Green functions
in the model Hamiltonians for the
three-dimensional higher-order topological insulators
defined in Eq.~(\ref{eq:3DHOT}).
For (a)$\Delta_2=0$, the system is a $\mathbb{Z}_2$
topological insulator and the system
becomes HOTI for (b) $\Delta_2=0.5$.
}
\label{fig:3DHOT}
\end{figure}

As shown in Fig.~\ref{fig:3DHOT},
the {traverses} of the zeros exist
even for the higher-order topological insulators.
Although $R_{4}$ lifts
the coincidences of the zeros and poles at some $k$ points such as X,
the coincidences still exist for $\Gamma$, M, and Z.
{This indicates that the band inversion still remains in the higher-order
topological insulators.}

\subsection{Higher-order topological insulators in two dimensions}
As an example of the 
two-dimensional higher-order topological insulators~\cite{Benalcazar_Science2017},
{we consider} the following Hamiltonian.
\begin{align}
H&= \notag \\
&\begin{pmatrix}
0                & 0                & ~iR_{z}+A        & iR_{x}+R_{y} \\
0                & 0                & ~iR_{x}-R_{y}    & -iR_{z}+A \\
-iR_{z}+A        & -iR_{x}-R_{y}    & ~0               & 0 \\
-iR_{x}+R_{y}    & iR_{z}+A         & ~0               & 0 
\end{pmatrix}
\label{eq:2DHOT}
\end{align}
where $R_x=E\sin{k_y}$,  $R_{y}=\gamma+E\cos{k_y}$, $R_{z}=E\sin{k_x}$, and
$A=\gamma+E\cos{k_x}$.
The eigenvalues are given as
\begin{align}
E_{0,\pm}&=\pm\sqrt{A^2+R_{x}^2+R_{y}^2+R_{z}^2}, \\
E_{1,\pm} &=\pm\sqrt{A^2+R_{x}^2+R_{y}^2+R_{z}^2}.
\end{align}
We note that the eigenvalues are doubly degenerate. 

Because the system has chiral symmetry,
{according to the procedure described above,
we define the following unitary matrix by taking $\vec{k}_{0}=(0,0)$.}

\begin{figure}[t!]
	\begin{center}
		\includegraphics[width=8cm]{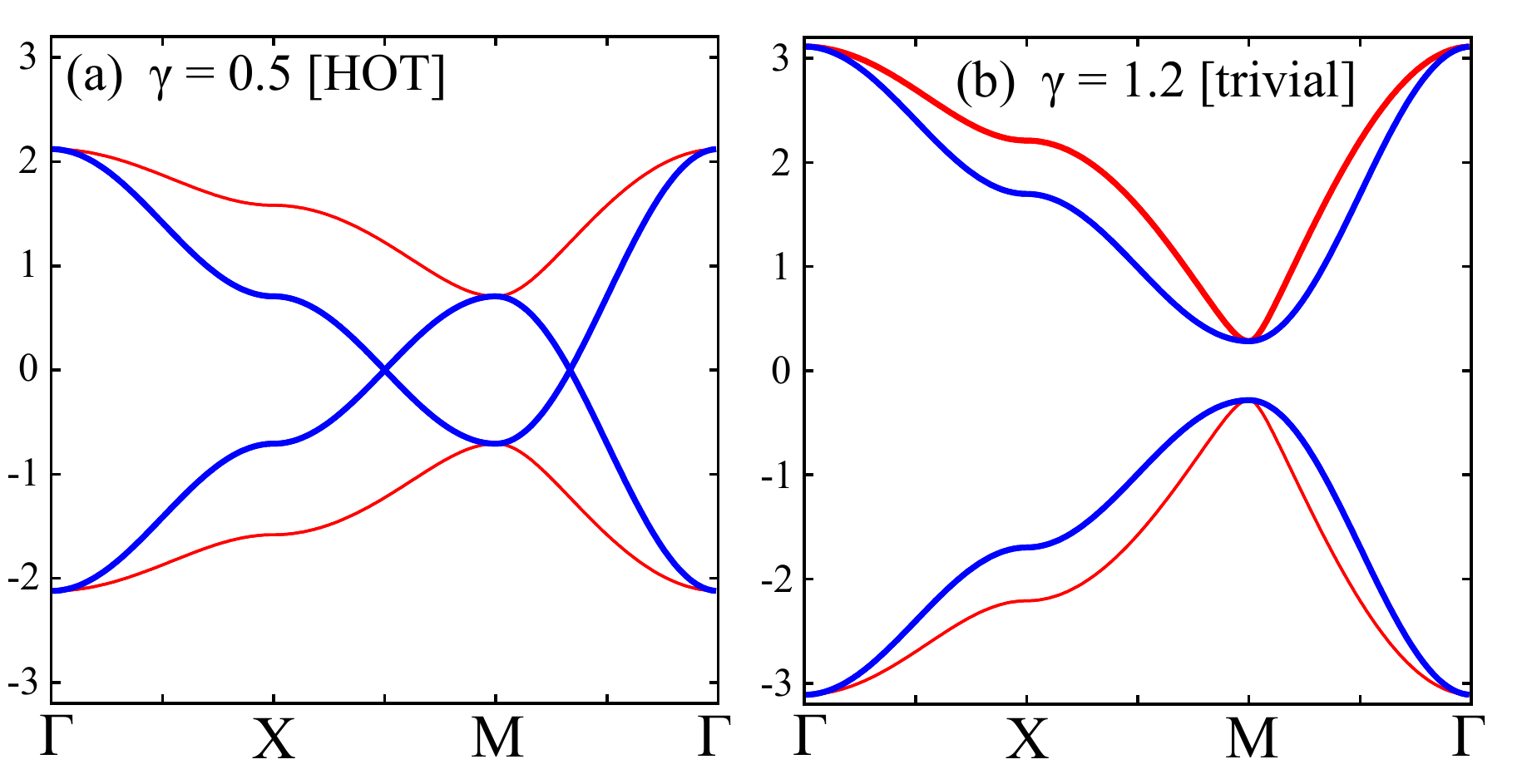}   
	\end{center}
\caption{~Band dispersions (thin red curves) 
and zeros (thick blue curves) of the Green functions
in the model for two-dimensional higher-order topological insulators
for (a) $\Gamma=0.5$ and (b) $\Gamma=1.2$.
}
\label{fig:2DHOT}
\end{figure}

{
We show zeros of the Green functions of transformed Hamiltonian 
$\tilde{H}=U^{\dagger}HU$ in Fig.~\ref{fig:2DHOT}.
We find that the zeros {traverse the band gap} in the higher-order topological 
insulator while they do not {traverse} in the trivial insulator.
This result indicates that the traverse of the zeros in the bulk system
is useful even for detecting the higher-order topological phases.}

\subsection{Higher-order topological insulator in the Fu-Kane-Mele model under a magnetic field}
Finally, we examine how the zeros of the Green functions
behave in the Fu-Kane-Mele model under a 
magnetic field~\cite{FuKaneMele_PRL2007,Fu_PRB2007,YiZhang_PRB2010}, which is defined as follows.
\begin{align}
R_{1}  &= 1+\delta t_{1}+[\cos{k_1}+\cos{k_2}+\cos{k_3}],  \\
R_{2}  &= \sin{k_1}+\sin{k_2}+\sin{k_3},  \\
R_{3}  &= \lambda_{\rm SO}[\sin{k_2}-\sin{k_3}-\sin{(k_2-k_1)}+\sin{(k_3-k_1)}],  \\
R_{4}  &= \lambda_{\rm SO}[\sin{k_3}-\sin{k_1}-\sin{(k_3-k_2)}+\sin{(k_1-k_2)}],  \\
R_{5}  &= \lambda_{\rm SO}[\sin{k_1}-\sin{k_2}-\sin{(k_1-k_3)}+\sin{(k_2-k_3)}],  \\
H      &=
\begin{pmatrix}
R_{5}+h_{z}   & ~R_{1}-iR_{2}  & R_{3}-iR_{4}  & ~0  \\
R_{1}+iR_{2}  & ~-R_{5}+h_{z}  & 0             & ~-R_{3}+iR_{4}   \\
R_{3}+iR_{4}  & ~0             & -R_{5}-h_{z}  & ~R_{1}-iR_{2}  \\
0             & ~R_{3}-iR_{4}  & R_{1}+iR_{2}  & ~R_{5}-h_{z} \\
\end{pmatrix},
\label{eq:FKQ}
\end{align}
where $k_{1}=(k_y+k_z)/2$,~$k_{2}=(k_x+k_z)/2$,~ $k_{3}=(k_x+k_y)/2$, and
$h_z$ represent the magnetic field.

It is shown that the three-dimensional $\mathbb{Z}_2$ topological
insulator appears in the Fu-Kane-Mele model.
The gapless surface
states are gapped out
when the magnetic field is applied because the magnetic 
field breaks the time-reversal symmetry.
Nevertheless, by analyzing the entanglement spectrum, 
Turner $et$ $al$. showed that the 
topological properties of the systems still remain~\cite{YiZhang_PRB2010}.
They demonstrated that the entanglement spectrum behaves as
gapless surface states, even under a magnetic field.
Recently, it has been established that the one-dimensional
hinge states appear in the $\mathbb{Z}_2$ topological insulator
under a magnetic field~\cite{Matsugatani_PRB2018,Fang_SA2019}. 
The higher-order topological nature 
may be the origin of the characteristic behavior in the entanglement spectrum.
{Here, we show} that the zeros of the Green functions
can also capture the 
topological properties of this system.

\begin{figure}[t!]
	\begin{center}
		\includegraphics[width=8cm]{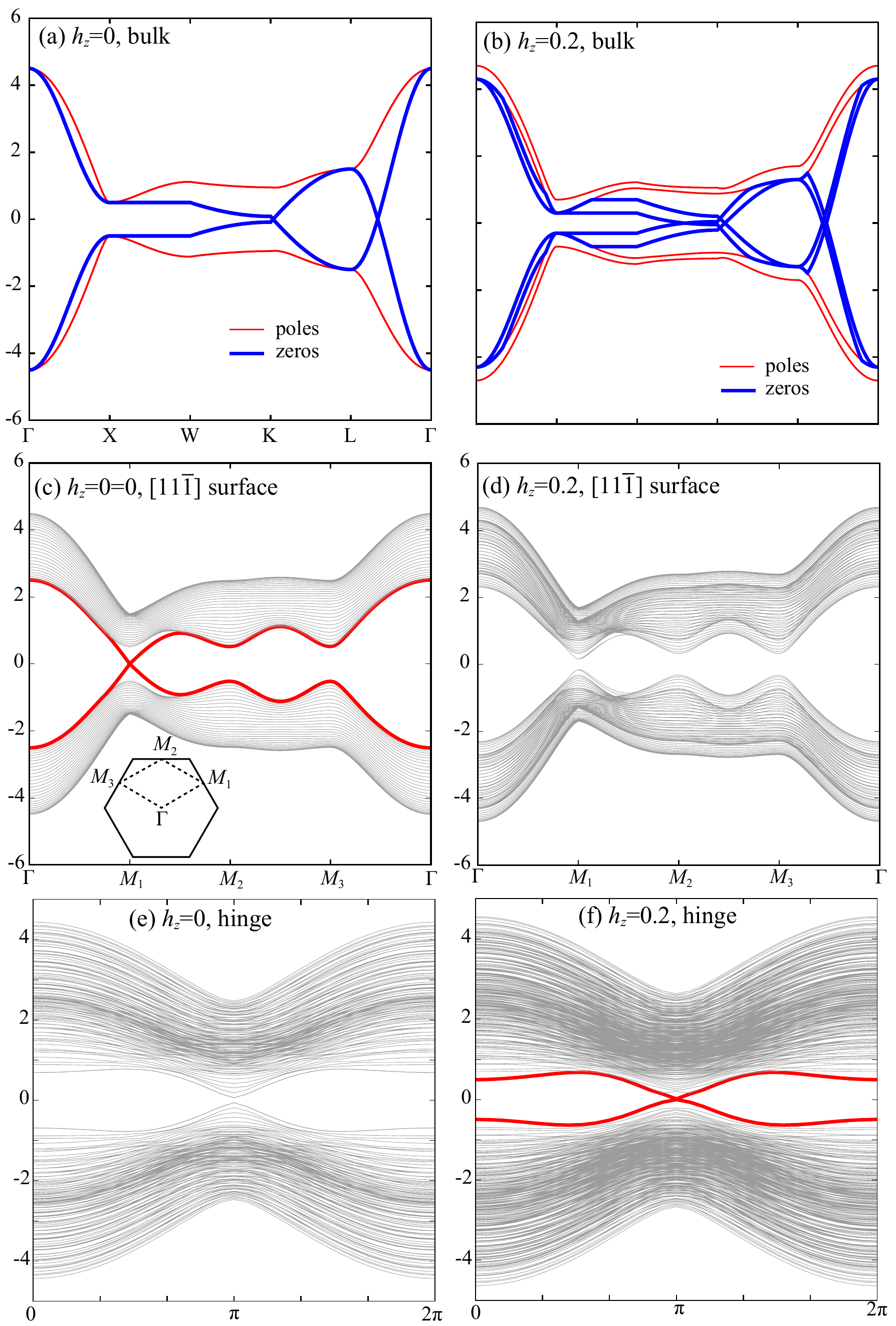}   
	\end{center}
\caption{
~Band dispersions (thin red curves) 
and zeros (thick blue curves) of the Green functions
in the Fu-Kane-Mele model 
for (a) $h_{z}=0$ and (b) $h_{z}=0.2$.
For this study, $\delta t=0.5$ and $\lambda_{\rm SO}=0.25$.
The {traverses} of the zeros survive even under finite magnetic fields.
~Band dispersions in the slab geometry ([11$\bar{1}$] surface) 
in the Fu-Kane-Mele model 
for (c) $h_{z}=0$ and (d) $h_{z}=0.2$.
Surface states (thick red curves shown in (c)) are 
gapped out by the magnetic field.
~Band dispersions in the hinge geometry 
in the Fu-Kane-Mele model 
for (e) $h_{z}=0$ and (f) $h_{z}=0.2$.
Gapless hinge states appear for $h_{z}=0.2$.
The unit vectors were employed as
$\vec{a}_{1}=(0,1,1)$, $\vec{a}_{2}=(1,0,1)$, and $\vec{a}_{3}=(1,1,0)$.
To realize the [11$\bar{1}$] surface (the one-dimensional hinge state), 
the periodicity was cut for $\vec{a}_{3}$  ($\vec{a}_{2}$ and $\vec{a}_{3}$).
The calculations of the surface and hinge states 
were performed using PythTB~\cite{PythTB}.
}
\label{fig:FKM}
\end{figure}

In Fig.~\ref{fig:FKM}(a) and (b),
the bulk band dispersions and 
the zeros of the Green functions for
$h_{z}=0$ and $h_{z}=0.2$, respectively, are presented.
For this investigation, $\delta t=0.5$ and $\lambda_{\rm SO}=0.25$.
{For visibility, we perform the unitary transformation that
consists of the eigenvectors at $\Gamma$ point.}
Even under the finite magnetic field,
it was determined that the {traverses} of the zeros 
of the Green functions can still survive.

To observe {the surface/edge states}, the two-dimensional 
surface states and the one-dimensional hinge states
were calculated in Fig.~\ref{fig:FKM}(c)-(f).
As shown in  Fig.~\ref{fig:FKM}(d),
the gapless surface states are gapped out by the magnetic field.
However, as shown in Fig.~\ref{fig:FKM}(f), 
the gapless one-dimensional edge (hinge) 
states appear under the magnetic field.
This result shows that a higher-order 
topological insulator is realized 
in the Fu-Kane-Mele model under the magnetic field and
is captured by the {traverses} of the zeros.

\section{Relation with edge states}
\label{sec:zeroedge}
The {traverse and the resultant cross} of the zeros resemble the gapless edge modes,
which appear in the topological insulators.
This section shows that the {traverse} of the zeros of the Green functions
has a relation with the gapless edge state, i.e., degeneracies of the
eigenvalues of the edge Hamiltonian guarantee
the existence of the zeros' surface in the band gap. 
We note that this argument is applicable to the Hamiltonians 
only with the nearest-neighbor hoppings. 
For the Hamiltonian with further-neighbor hoppings, 
the argument can not be directly applied.

We consider the two-dimensional Chern insulators as an example,
which is defined in Eq.~(\ref{eq:Chern}).
We rewrite the Hamiltonians in the
real space for the $x$ direction (the length is set to $L$) as
\begin{align}
&H_{L}(p)=
\begin{pmatrix}
h_{0}          ~~& h_{x}    ~~& 0       ~~& \ldots  ~~&0      ~~& p\times h_{-x} \\
h_{-x}         ~~& h_{0}    ~~& h_{x}   ~~& 0       ~~&0      ~~& 0             \\
0              ~~& h_{-x}   ~~& h_{0}   ~~& h_{x}   ~~&0      ~~& 0             \\
\vdots         ~~& \vdots   ~~& \ddots  ~~& \ddots  ~~&\ddots ~~& \vdots        \\
0              ~~& 0        ~~& 0       ~~& h_{-x}  ~~&h_{0}  ~~& h_{x}    \\
p\times h_{x}        ~~& 0        ~~& 0       ~~& 0       ~~&h_{-x} ~~& h_{0}    
\end{pmatrix} \label{eq:bulk}, \\
&h_{0}=
\begin{pmatrix}
2+m-\cos{k_{y}}  ~~& -i\sin{k_{y}} \\
i\sin{k_{y}}     ~~& -(2+m)+\cos{k_{y}}  
\end{pmatrix},\\
&h_{x}=
\begin{pmatrix}
1/2  ~~& i/2 \\
i/2  ~~& -1/2 \\
\end{pmatrix},\\
&h_{-x}=
\begin{pmatrix}
1/2   ~~& -i/2 \\
-i/2  ~~& -1/2  
\end{pmatrix},
\end{align}
where $p$ is the scholar value and controls the boundary conditions.
For example, the open-boundary condition corresponds to
$p=0$, and the periodic-boundary condition corresponds to $p=1$.
We note that the periodic Hamiltonian ($p=1$) includes the $L-1$ edge Hamiltonian
\begin{align}
H_L(p=1)=
\begin{pmatrix}
h_{0}       ~~~~& B             \\
B^{\dagger} ~~~~& H_{L-1}(p=0)
\end{pmatrix},
\end{align}
where $B=(h_{x},0,\ldots,0,p\times h_{-x})$.
For simplicity, we denote $H_{\rm real}(k_{y})=H_{L}(p=1,k_{y})$
and $H_{\rm edge}(k_{y})=H_{L-1}(p=0,k_{y})$.
{We note the relation between the edge states and the minor matrix 
are studied in the context of the Hermiticity of the tight-binding Hamiltonian
operators~\cite{Fukui_PRR2020}.}

Here, we define
the partial Fourier transformed Green function as
\begin{align}
\bar{G}_{n}(k_y,\omega) = \int_{-\pi}^{\pi} G_{n}(k_x,k_y,\omega)d k_x.
\label{eq:partialF}
\end{align}
We note that $\bar{G}_{n}$ is the Green functions of $H_{\rm real}(k_{y})$, i.e.,
\begin{align}
\bar{G}_{n}(k_y,\omega) = [\omega-H_{\rm real}(k_y)]^{-1}.
\end{align}
From this definition,
if $\bar{G}_{n}(k_y,\omega)$ becomes zero,
the following relation is satisfied.
\begin{align}
\bar{G}_{n}(k_y,\omega)=\int_{-\pi}^{\pi}G_{n}(k_x,k_y,\omega)d k_x = 0.
\label{eq:intG}
\end{align}
This indicates that the $G_{n}(k_x,k_y,\omega)$
changes its sign as a function of $k_{x}$.

The zeros of $\bar{G}_{n}(k_y,\omega)$ 
are given by the eigenvalues of the minor matrix $M_{n}$ of $H_{\rm real}(k_{y})$.
Based on the periodicity, we only consider the eigenvalues
of $M_{0}$ and $M_{1}$. 
Because $M_{0}$ and $M_{1}$ include the edge Hamiltonian, 
$H_{\rm edge}$ can be regarded as a minor matrix of $M_{0}$ and $M_{1}$.
Thus, from the Cauchy interlacing identity (see Appendix E and F),
the eigenvalues of $H_{\rm edge}$ are located between the eigenvalues of $M_{0}$ and $M_{1}$ and
vice versa. This relation can be expressed as
\begin{align}
&E_{i-1}(H_{\rm edge}) \leq E_{i}(M_{0})=\bar{\zeta}_{0}(H_{\rm real}) \leq E_{i}(H_{\rm edge}), \notag \\
&E_{i-1}(H_{\rm edge}) \leq E_{i}(M_{1})=\bar{\zeta}_{1}(H_{\rm real}) \leq E_{i}(H_{\rm edge}).
\end{align}
If the eigenvalues of the edge Hamiltonian are doubly degenerate,
the zeros of the bulk Hamiltonian should coincide, i.e.,
\begin{align}
\bar{\zeta}_{0}(H_{\rm real}) = \bar{\zeta}_{1}(H_{\rm real}).
\end{align}
This indicates that the gapless edge states (they cross in the bandgap)
inevitably induce the degeneracy of the zeros in $\bar{G}_{n}$.

Using numerical calculations, 
we demonstrate the application of this argument to the Chern insulators.
In Fig.~\ref{fig:Edge}, 
the poles of the zeros of the Green functions
for the periodic Hamiltonian $H_{L=20}(p=1)$ are defined in
Eq.~(\ref{eq:bulk}) for $m=-0.5$.
We also show the poles of the edge Hamiltonian (edge states) 
for $H_{L=19}(p=0)$.
We can confirm that
the zeros of the periodic Hamiltonian exist between the
poles of the edge Hamiltonian, and
they degenerate when the poles of the edge Hamiltonian cross ($k_{y}=0$).

In general, for the $m$-orbital system (the size of $h_0$ is $m\times m$),
it can be shown that 
the $m$-fold degeneracy in the edge states (gapless edge states) 
{induces} the $m$-fold degeneracy 
in the zeros of $\bar{G}$.
The proof of this statement 
is shown in Appendix \ref{appendixB} and \ref{appendixC}.
However, because the proof is based on the Cauchy interlacing inequality,
it can not be applied to the Hamiltonian with the further-neighbor hoppings, where
degeneracy of the edge states ($m_{\rm edge}$) is less 
than the number of orbitals ($m_{\rm orb}$), i.e., $m_{\rm edge}<m_{\rm orb}$. 
For example, if we introduce the next-nearest neighbor hopping, the size of $h_{0}$
becomes $4\times4$ while the degeneracy of the edge states still remains $2$.
In this situation, the Cauchy interlacing inequality can not be applied
and the degeneracy of the edge states does not necessarily induce
the degeneracy in the zeros of $\bar{G}$.

Here, we argue how the {traverses} of the zeros in $\bar{G}$
are related with the zeros of the Green functions in the momentum space.
We first consider the region where the $\omega$ is negative ($\omega<0$)
and the 0th component of the Green function ($\bar{G}_{0}$ and $G_{0}$).
For $\omega<0$, at the certain $k_{y}=\pm k_{0}$, $\bar{G}_0(k_y,\omega)$ becomes 0.
This indicates that the finite positive and negative regions exist in the
original Green functions at $\pm k_{0}$ according to Eq.~(\ref{eq:intG}).
Since the positive and negative regions do not immediately vanish even 
if we adiabatically change $k_y$, the zero's surface exists
in the original Green functions $G_{0}$ as shown in Fig.~\ref{fig:Edge}(a).

At $\omega=0$,  $\bar{G}_0(k_y,\omega)$ becomes 0 at $k_{y}=0$. 
This also indicates the existence of the zero's surface around $k_{y}=0$
(see Fig.~\ref{fig:Edge} (b)).
We also note that the zero's surface can not immediately vanish even if we 
change $\omega$, the zero's surface survives for $\omega>0$ as is
shown Fig.~\ref{fig:Edge} (c).
From this consideration, we can say that the zero's surface of $G_{0}$
exists at least for $\omega\leq0$

The opposite thing occurs for $G_1$ and $\bar{G}_1$, i.e.,
the zero's surface exists at least for $\omega\geq0$.
These results indicate that the zeros' surfaces exist in the band gap
if the gapless edge states exist. This means that
at least one diagonal component of the Green functions becomes zero
in the band gap due to the existence of the gapless edge states.
The existence of the zeros' surface is consistent with the
existence of the traverses 
of the zeros in the bulk Green functions.
We note that, however, 
the traverses of the zeros are {\it not} guaranteed
by the degeneracy of the zeros of $\bar{G}$.

\begin{figure}[t!]
	\begin{center}
		\includegraphics[width=8cm]{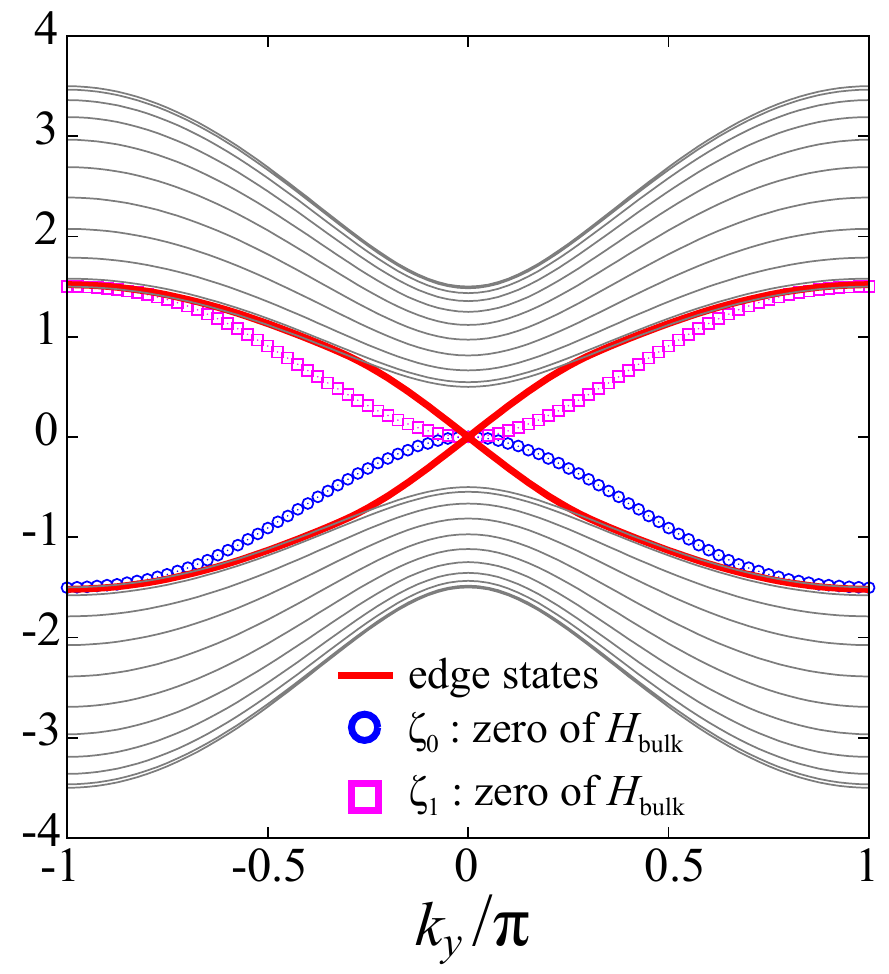}   
	\end{center}
\caption{~Band dispersions (thin black curves) 
and zeros of the Green functions (blue circles and magenta squares)
for the bulk Hamiltonian $H_{\rm reak}(k_{y})$. 
For this investigation, $L=20$ and $m=-0.5$.
The zeros of the Green function are plotted in the bandgap.
It is also established that the eigenvalues (thick red curves)
of $H_{\rm edge}(k_{y})$ are in the bandgap.
As described in the main text, the cross 
of the eigenvalues of  $H_{\rm edge}(k_{y})$
{induces} the cross of the zeros {in the partially Fourier transformed Green functions $\bar{G}_{0}(k_y,\omega)$}.
}
\label{fig:Edge}
\end{figure}%

\begin{figure}[t!]
	\begin{center}
		\includegraphics[width=8cm]{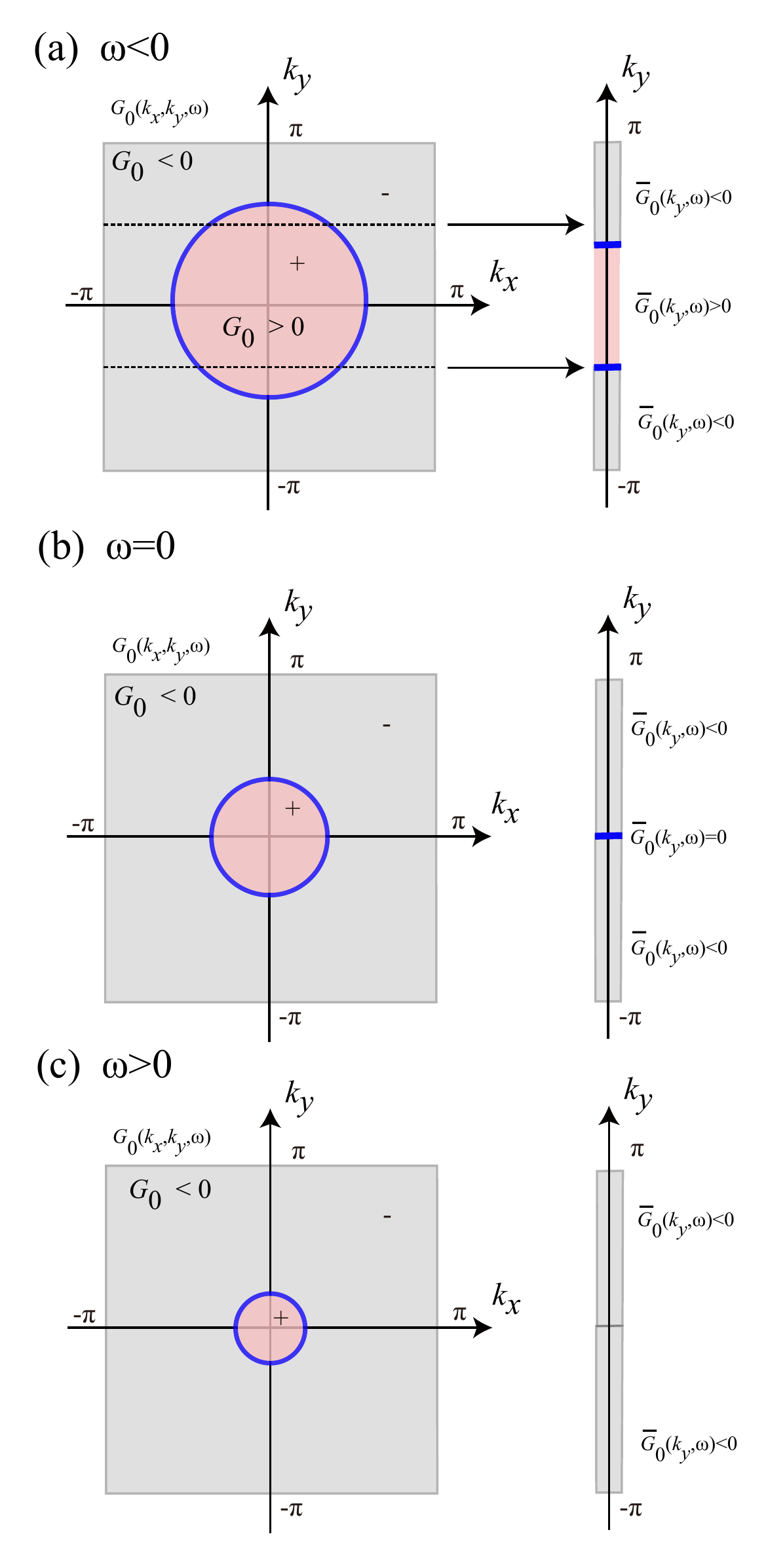}   
	\end{center}
\caption{~Sign of the Green functions of $G_{0}(k_x,k_y,\omega)$ (left panel) and
the partially Fourier transformed Green functions ($\bar{G}_{0}(k_y,\omega)$) (right panel)
for (a) $\omega>0$, (b) $\omega=0$, and (c) $\omega<0$.
The zeros of the Green functions are denoted by the thick (blue) lines.
}
\label{fig:Edge}
\end{figure}%

\section{Summary} 
In summary, this study focused on the zeros of the diagonal components of the 
Green functions in topological insulators.
Based on the arguments that were used in the eigenvector-eigenvalue identity~\cite{Denton_BAMS2022},
it was first demonstrated that the zeros of the
diagonal components of the Green functions are given by the eigenvalues
of the minor matrix $M_{n}$, which is obtained
by removing the $n$th row and column from the
original Hamiltonian.
This mathematical foundation offers an efficient way to study
the zeros of the Green functions both analytically and numerically.
{We have also shown that the zeros can visualize the information of the band inversions via
the eigenvector-eigenvalue identity.}

For a two-dimensional Chern insulator,
which is a canonical model of 
two-dimensional a topological insulator,
it is established that the traverse of the zeros 
is {a} key to distinguishing 
topological insulators from trivial insulators. 
For the Chern insulators, this study has explicitly shown 
that the existence of the Chern number guarantees the traverse of the zeros
in bulk systems. 
It has been demonstrated that the traverse of the zeros is gauge invariant,
although the positions of the zeros are not.
It is observed that the traverses of the zeros
can occur in the other class of the topological insulators
without chiral symmetries such as
the $\mathbb{Z}_2$ (class AII) topological insulators in
two and three dimensions and
$2\mathbb{Z}$ (class AI) topological insulators in four dimensions.

{
We give a general argument that the band inversions
induce the traverses of the zeros in the band gap.
Since the band inversions occur in topological phases,
the traverses of the zeros universally occur in the topological phase.
This argument also shows that the traverse of the zeros always occur
if we take the proper unitary transformation. The unitary transformation
is used for identifying the topological phase with chiral symmetry.}

For topological insulators with chiral symmetry,
this study demonstrated that the zeros also {traverse} the bandgap
{when we take the suitable unitary transformation defined in Eq.~(\ref{eq:unitary}). 
By taking the model Hamiltonians for class BDI, AIII, and CII topological
insulators, this investigation has shown that the {traverses} of the zeros always 
occur in topological phases.

Since the {traverses} of the zeros are guaranteed by
the existence of {the band inversions} 
it is expected that this is useful for detecting other
exotic topological phases, which are not listed in the
conventional periodic table for topological insulators 
(Table ~\ref{PeriodicTable}). 
As an example of these exotic topological phases,
we consider the higher-order topological insulators. The {traverse} of the zeros
is also useful for detecting 
the higher-order topological insulators in two and three dimensions.

{We also show that the gapless edge states guarantee
the existence of the zeros' surface in the band gap for
the Hamiltonian with the nearest neighbor hoppings.
Although the relation with the edge states and the 
{behavior} of the zeros in the bulk system can be proved only for the limited case,
this result implies that the zeros have close relation with
the edge states and offers another point of view of 
the bulk edge-correspondence~\cite{HatsugaiBulkEdge}.}

The comprehensive analysis in this study demonstrates that
the zeros of the Green functions
can be used as a simple visual detection tool
for topological phases {via the band inversions}. 
It may be useful when searching for new topological phases 
because the {traverses} of the zeros can be easily detected
without any assumptions.
In addition, since this method does not require full gap states but only needs
the {traverse} of the zeros, the zeros of the Green functions can be useful to identify the
exotic 
topological semimetals
such as the Weyl~\cite{Shindou_PRL2001,Murakami_NJ2007,Wan_PRB_2011,Burkov_AR2018} 
and the Topological Dirac semimetals~\cite{Wang_PRB2012,Morimoto_RPB2014,Yang_NCom2014}.

{It needs to be noted that the accidental crosses of the zeros
in the trivial phases may occur, as demonstrated for the CII topological insulator.}
We want to emphasize that this type of cross
is {\it not} protected by the traverse of the zeros, i.e., the band inversion.
By visualizing the behavior of the zeros in the band gap,
it is easy to distinguish whether 
the crosses of the zero are accidental or not.
{We also note that the accidental crosses 
appear in the Rice-Mele model~\cite{Rice_RPL1982,Xiao_RMP2010},
which has non-topological edge states.
In the Rice-Mele model, since the zeros do not traverse the band gap either,
we can distinguish whether the crosses are accidental.}
Thus, the {traverse} of the zeros of the Green functions is 
useful when searching for the topological phases since
the absence of the {traverses} indicates that the 
system is not topological in the sense that they have no band inversions.
This fact may be useful in the screening process 
for searching the topological materials 
by combining with the high-throughput $ab$ $initio$ 
calculations~\cite{Zhang_Nature2019,Tang_Nature2019,Vergniory_Nature2019,Xu_2020Nature}.
}

{
In this paper, the rediscovered 
eigenvector-eigenvalue identity is essentially used for showing the
existence of the traverse of the zeros due to the band inversion.
Since the band inversion universally occurs in the topological phases,
the traverse of the zeros offers a useful guideline for
identifying the topological phases.
Thus, our paper gives a direct application of 
the eigenvector-eigenvalue identity 
in condensed matter physics, 
which is a fundamental relation in linear algebra but 
has long been overlooked.
}

This study was restricted to the
analysis of topological insulators for simplicity; however,
a similar analysis is possible for topological superconductors,
{where the band inversion also occurs}.
{It is known that several topological superconductors can be mapped to the 
models of the topological insulators,e.g., the model for the 
class D topological superconductors in the two dimension 
is mapper to the model for the Chern insulator~\cite{Wang_PRB2021}.
In those systems, 
the traverses zeros also appear in the topological superconductors.}
Furthermore, a similar analysis might be possible 
for the disordered~{\cite{Li_PRL2009,Kobayashi_PRL2013}} 
and interacting~{\cite{Gurarie,LeiWang_PRB2012,WWK_AR2014}} systems
because the Green functions 
for the correlated and/or disordered systems are well-defined.
We note that zeros of the Green functions in correlated electron systems are recently studied using dynamical mean-field theory~\cite{Tran_PRB2022}.
These unexplored issues are intriguing but need to be further investigated.

\begin{acknowledgements}
This work was supported by JSPS KAKENHI. 
(Grant Nos.~{JP16H06345, JP19K03739, and JP20H01850}).
{This work was also supported by the 
National Natural Science Foundation of China (Grant No. 12150610462).}
{T.M. thanks Masafumi Udagwa, Yutaka Akagi, and Satoru Hayami 
for fruitful discussions in the early stage of this study.}
{T.M. also thanks Tomi Ohtsuki for useful discussions.}
T.M. would also like to thank Yi Zhang for the valuable discussion 
on Ref.~[\onlinecite{YiZhang_PRB2010}].
T.M. was also supported by the Building of Consortia for 
the Development of Human Resources 
in Science and Technology from the MEXT of Japan.
\end{acknowledgements}

\appendix
\section{Proof of the Cauchy interlacing inequalities}
\label{ap:Cauchy}
Although the proof of the Cauchy interlacing
inequalities is shown in the literature~\cite{Wilkinson_1963,Hwang_2004},
we have provided another proof of the inequalities based on the
structure of the Green functions
to make this paper self-contained.
To prove the Cauchy interlacing inequalities, 
the definition of the Green function (Eq.~(\ref{eq:green})) is rewritten as
\begin{align}
G_{n}(\boldsymbol{k},\omega)&=\frac{N_{n}(\boldsymbol{k},\omega)}{\prod_{i}{(\omega-E_{i})}}, \label{eq:GAll}\\
N_{n}(\boldsymbol{k},\omega)&=\sum_{j}|\Psi_{j}^{(n)}|^{2}\prod_{i\neq j}(\omega-E_{i}). \label{eq:Numerator}
\end{align}
The roots of the ($n-1$)th polynomial $N_{n}(\boldsymbol{k},\omega)$ correspond to
the zeros of $G_{n}(\boldsymbol{k},\omega)$, i.e., 
the eigenvalues of the minor matrix $M_n({\vec{k}})$.

From Eq.~(\ref{eq:Numerator}), it can be shown that 
$N_{n}(\vec{k},\omega)$ changes its sign between $E_{i}(\vec{k})$ and $E_{i+1}(\vec{k})$, i.e., 
$N_{n}(\boldsymbol{k},E_{i}(\boldsymbol{k}))\cdot N_{n}(\boldsymbol{k},E_{i+1}(\boldsymbol{k}))<0$ 
unless $|\Psi^{(n)}_{i}|=0$ or $|\Psi^{(n)}_{i+1}|=0$. 
This indicates that the $N_{n}(\vec{k},\omega)$ has at least one real root between
$E_{i}(\vec{k})$ and $E_{i+1}(\vec{k})$.
Because the $N_{n}(\vec{k},\omega)$ 
is the ($n-1$)th polynomial with respect to $\omega$, 
the number of roots {between $E_{i}(\boldsymbol{k})$ and $E_{i+1}(\boldsymbol{k})$} should be one.
For $|\Psi^{(n)}_{i}(\vec{k})|=0$, 
$\omega=E_{i}(\vec{k})$ is a 
root of $N_{n}(\vec{k},\omega)$.
This is consistent 
with the eigenvector-eigenvalue identity.

Thus, for both $|\Psi_{i}^{(n)}|=0$ and $|\Psi_{i}^{(n)}|\neq0$, 
$N_{n}(\vec{k},\omega)$ has one real 
root between $E_{i}(\boldsymbol{k})$ and $E_{i+1}(\boldsymbol{k})$.
In other words, one zero of $G_{n}(\boldsymbol{k},\omega)$ should exist
between every adjacent eigenvalue (poles of the Green function). 
This was also pointed out in a previous study~~\cite{Seki_PRB2017}.
This represents the
Cauchy interlacing inequalities.

\section{Relation between the band inversion and the topological invariants}
\label{ap:bandinversion}
Let $\vec{e}_{i}(\vec{k})$ is the $i$th eigenvectors of Hamiltonian $H(\vec{k})$.
We take $N$ ($M$) as the number of the occupied  (unoccupied) states.
We define the overlap matrix $U$ as 
\begin{align}
U(\vec{k}_{0},\vec{k}_{1}) = \mathcal{E}(\vec{k}_{0})^{\dagger}\mathcal{E}(\vec{k}_{1}),
\end{align}
{where $\mathcal{E}(\vec{k})$ is the set of the eigenvectors at $\vec{k}$,}
\begin{align}
{\mathcal{E}(\vec{k}) = [\vec{e}_{0}(\vec{k}),\vec{e}_{1}(\vec{k}),\dots,\vec{e}_{N+M-1}(\vec{k})]}.
\end{align}
{Since $\mathcal{E}(\vec{k})$ is unitary,
$U$, which is a product of two unitary matrices,  is also unitary.} 
For the unitary matrix $U$,
we consider the following block matrix representation:
\begin{align}
U=
\begin{pmatrix}
U_{11} ~~& U_{12} \\
U_{21} ~~& U_{22}
\end{pmatrix},
\end{align}
{where
$U_{11}$ ($U_{22}$) is the overlap matrix for the occupied (unoccupied) states
and it is the $N\times N$  ($M\times M$) unitary matrix
while $U_{12}$ ($U_{21}$) is the $N\times M$ ($M \times N$) overlap matrix between the occupied and unoccupied states.}
We note that {$\det{U_{11}}$}
is a key
{quantity} 
for identifying
{the non-trivial
Chern and $Z_{2}$ topological insulators}.
{When a system is the non-trivial Chern or $Z_2$ topological insulator,
there exists, at least, a single pair of $\vec{k}$, $(\vec{k}_0,\vec{k}_1)$, that satisfies $\det{U_{11}}=0$.}

For the Chern insulator, it is shown that, 
{if $\det{U_{11}}=0$ holds at a pair of momenta, $(\vec{k}_0,\vec{k}_1)$, irrespective of the choice of the gauge,
the gauge fixing is impossible~\cite{Hatsugai_JPSJ2004}.
Thus, if $\det{U_{11}}=0$ holds at a pair of momenta, there is a non-trivial Chern number.}

In the $Z_{2}$ topological insulators~\cite{KaneMele,Fukui_JPSJ2007},
it is also shown that the 
non-trivial $Z_{2}$ topological invariant induces
zeros of the Pfaffian of the following overlap matrix 
\begin{align} 
P(\vec{k})={\rm Pf}[\mathcal{E}_{\rm occ}(\vec{k})^{\dagger}(\vec{k})\Theta\mathcal{E}_{\rm occ}(\vec{k})],
\end{align} 
where $\Theta$ is the time-reversal operator {and $\mathcal{E}_{\rm occ}(\vec{k})=
[\vec{e}_{0}(\vec{k}),\vec{e}_{1}(\vec{k}),\dots,\vec{e}_{N-1}(\vec{k})]$}.
Since $\Theta\mathcal{E}_{\rm occ}(\vec{k})$ can be obtained from {an} 
unitary transformation of $\mathcal{E}(-\vec{k})$,
we obtain
\begin{align} 
\Theta\mathcal{E}_{\rm occ}(\vec{k})=\mathcal{E}_{\rm occ}(-\vec{k})W,
\end{align} 
{and} 
\begin{align} 
&P(\vec{k})={\rm Pf}[\mathcal{E}_{\rm occ}(\vec{k})^{\dagger}\mathcal{E}_{\rm occ}(-\vec{k})W] \notag\\
&={\rm Pf}[\mathcal{E}_{\rm occ}(\vec{k})^{\dagger}\mathcal{E}_{\rm occ}(-\vec{k})]\times{\rm Pf}[W],
\end{align} 
where $W$ is the unitary matrix.
{Then}, we obtain
\begin{align} 
|\det{U}_{11}(\vec{k},-\vec{k})|=|P(\vec{k})|^2.
\end{align} 
{Therefore,} $|P(\vec{k})|=0$ is equivalent
to $|\det{U_{11}(\vec{k},-\vec{k})}|=0$.

When the  band inversion occurs between 
$\vec{k}_{0}$ and $\vec{k}_{1}$, i.e.,
{for all $i\in [0,N-1]$,
$n_{0}\in\rm [0,N-1]$ exists}
such that
$\vec{e}_{i}(\vec{k}_{0})^{\dagger}\vec{e}_{n_{0}}(\vec{k}_{1})=0$. 
This means that one column of $U_{11}$ is zero and{, thus,} $\det{U_{11}}=0$.
{In the following,}
we {will} show that {the} opposite statement also holds, i.e.,
$\det{U_{11}}=0$ induces the band inversion.

First, we notice that, for 
{$U_{11}, U_{12}, U_{21}$, and $U_{22}$}, 
the following relations hold~\cite{Chen_arXiv2019}: 
\begin{align}
&{U_{11}U_{11}^{\dagger}}={I-U_{12}U_{12}^{\dagger}},\\
&{U_{22}^{\dagger}U_{22}}={I-U_{12}^{\dagger}U_{12}},\\
&|\det{U_{11}}|^2=|\det{U_{22}}|^2.
\end{align}
We note that from $|\det{U_{11}}|^2=|\det{U_{22}}|^2$,
if the occupied bands are non-trivial~($|\det{U_{11}}|=0$), the unoccupied bands
are also non-trivial~($|\det{U_{22}}|=0$). 

If $\det{U_{11}}=\det{U_{22}}=0$, $U_{11}$ and $U_{22}$ have the 
following {singular value decompositions}:
\begin{align}
&U_{11}= R\Sigma_{11}S^{\dagger}, \\
&\Sigma_{11}={\rm diag}[0,\lambda_{1}^{(1)},\lambda_{1}^{(2)}, \dots, \lambda_{1}^{(N-1)}], \\
&U_{22}= T\Sigma_{22}V^{\dagger},\\
&\Sigma_{22}={\rm diag}[0,\lambda_{2}^{(1)},\lambda_{2}^{(2)}, \dots, \lambda_{2}^{(M-1)}]
\end{align}
where $R,S,T$ and $V$ are unitary matrices, and $[0,\lambda_{1}^{(1)},\lambda_{1}^{(2)}, \dots, \lambda_{1}^{(N-1)}]$
($[0,\lambda_{2}^{(1)},\lambda_{2}^{(2)}, \dots, \lambda_{2}^{(M-1)}]$) are the singular values of $U_{11}$ ($U_{22}$).
{Here, we explicitly denote that the smallest singular value is zero from the assumption, $\det{U_{11}}=\det{U_{22}}=0$.}
Thus, we obtain
\begin{align}
&U_{11}U_{11}^{\dagger}=RD_{11}R^{\dagger},\\
&D_{11}={\rm diag}{[0,|\lambda_{1}^{(1)}|^2,\dots]},\\
&U_{22}^{\dagger}U_{22}=VD_{22}V^{\dagger},\\
&D_{22}={\rm diag}{[0,|\lambda_{2}^{(1)}|^2,\dots]}.
\end{align}
Since $U_{11}U_{11}^{\dagger}$ ($U_{22}^{\dagger}U_{22}$) is equivalent to
$I-U_{12}U_{12}^{\dagger}$ ($I-U_{12}^{\dagger}U_{12}$), 
we can simultaneously diagonalize using the same unitary matrix
\begin{align}
&R^{\dagger}U_{11}U_{11}^{\dagger}R=R^{\dagger}(I-U_{12}U_{12}^{\dagger})R=D_{11}, \\
&V^{\dagger}U_{22}^{\dagger}U_{22}V=V^{\dagger}(I-U_{12}^{\dagger}U_{12})V=D_{22}.
\end{align}
This indicates that we can take
\begin{align}
&U_{12}=R\Sigma_{12}V^{\dagger},
\end{align}
where the diagonal elements of $\Sigma_{12}$  are given by
$[1,\sigma_{1}^{(1)},\dots]$.
Similarly, we obtain
\begin{align}
&U_{21}=T\Sigma_{21}S^{\dagger},
\end{align}
where the diagonal elements of $\Sigma_{21}$  are given by
$[1,\sigma_{2}^{(1)},\dots]$.

This means that we have the following decomposition of $U$ as follows.
\begin{align}
&U=
\begin{pmatrix}
U_{11} ~~& U_{12} \\
U_{21} ~~& U_{22}
\end{pmatrix}=
\begin{pmatrix}
R~~& 0 \\
0 ~~& T
\end{pmatrix}
\begin{pmatrix}
\Sigma_{11}~~& \Sigma_{12} \\
\Sigma_{21}~~& \Sigma_{22}
\end{pmatrix}
\begin{pmatrix}
S^{\dagger}~~& 0 \\
0 ~~& V^{\dagger}
\end{pmatrix}\notag\\
&=\begin{pmatrix}
R~~& 0 \\
0 ~~& T
\end{pmatrix}
\begin{pmatrix}
0         & 0                 &\dots  ~&1      ~&~0                 &\dots \\
0         & \lambda_{1}^{(1)} &\dots  ~&0      ~&~\sigma_{1}^{(1)}  &\dots \\
\vdots    & \vdots            &\ddots ~&\vdots ~& \vdots            &\ddots\\
1~~       & 0~~               &\dots  ~&0      ~& 0                 &\dots \\
0~~       & \sigma_{2}^{(1)}  &\dots  ~&0      ~& \lambda_{2}^{(1)} &\dots \\
\vdots    & \vdots            &\ddots ~&\vdots ~& \vdots            &\ddots 
\end{pmatrix}
\begin{pmatrix}
S^{\dagger} ~~& 0 \\
0 ~~& V^{\dagger}
\end{pmatrix}.
\end{align}
By defining
\begin{align}
&J=
\begin{pmatrix}
R ~~& 0 \\
0 ~~& T
\end{pmatrix},
\Sigma=
\begin{pmatrix}
\Sigma_{11}     ~~& \Sigma_{12} \\
\Sigma_{21}~~& \Sigma_{22}
\end{pmatrix},
K^{\dagger}=
\begin{pmatrix}
S^{\dagger} ~~& 0 \\
0 ~~& V^{\dagger}
\end{pmatrix},
\end{align}
we can rewtire $U(\vec{k}_{0},\vec{k}_{1})$ as
\begin{align}
\mathcal{E}(\vec{k}_{0})^{\dagger} \mathcal{E}(\vec{k}_{1})
= U(\vec{k}_{0},\vec{k}_{1})
= J\Sigma K^{\dagger},
\end{align}
while
\begin{align}
[\mathcal{E}(\vec{k}_{0})J]^{\dagger}[\mathcal{E}(\vec{k}_{1})K]=\Sigma.
\notag
\end{align}
Since  $K$ and $J$
act on the occupied and unoccupied bands separately,
each occupied eigenstate in $\tilde{\mathcal{E}}(\vec{k_{0}})=\mathcal{E}(\vec{k}_{0})J$ 
and $\tilde{\mathcal{E}}(\vec{k_{1}})=\mathcal{E}(\vec{k}_{1})K$  only consists
the occupied states in the original basis.
From the structure of $\Sigma_{11}$, i.e. one column of $\Sigma_{11}$ is $0$,
we can show that
occupied bands at $\vec{k}_{0}$ 
does not include at least one occupied state at $\vec{k}_{1}$.
We take one missing occupied eigenstates as $\vec{e}_{n_{0}}(\vec{k}_{1})$. 
Thus, we obtain
\begin{align}
U_{i,n_{0}} = 
\vec{e}_{i}^{\dagger}(\vec{k}_{0})\vec{e}_{n_{0}}(\vec{k}_{1}) = 0 ~~~~~~~{{i\in[0,N-1]}}.
\end{align}
This shows that $\vec{e}_{n_{0}}(\vec{k}_{1})$
only consists of unoccupied states at $\vec{k}_{0}$.
This proves $\det{U_{11}}=0$ induces
the band inversion.

\section{Condition for the traverse of the zero}
\label{ap:traverse}
As we explained in main text,
when $\vec{e}_{0}(\vec{k}_0)$
is orthogonal to $\vec{e}_{0}(\vec{k}_1)$,
$\tilde{\zeta}_{0}(\vec{k}_1)$ coincides with $E_{0}(\vec{k}_1)$. 
However, this orthogonal relation does not always 
mean the traverse of $\tilde{\zeta}_{0}$.
When $\tilde{\zeta}_{0}$ approaches
to $E_{0}(\vec{k}_1)$ from the upper side of $E_{0}(\vec{k}_1)$
($\lim_{\vec{k}\rightarrow\vec{k}_{1}}\tilde{\zeta}_{0}(\vec{k})=E_{0}(\vec{k}_1)+0$), 
 $\tilde{\zeta}_{0}$ traverses the band gap. 
In this appendix, we give the condition when the 
$\tilde{\zeta}_{0}$ traverses the band gap.

We consider the point where the momentum is slightly
different from $\vec{k}_{1}$, i.e., $\vec{k}_{1}+\Delta\vec{k}$.
The position where $\tilde{G}_{0}(\vec{k}_{1}+\Delta\vec{k},E_{0}(\vec{k}_1)+\delta)=0$
is given by $\tilde{N}_{0}(\vec{k}_{1}+\Delta\vec{k},E_{0}(\vec{k}_1)+\delta)=0$,
where $\tilde{N}_{0}$ is defined in Eq.(\ref{eq:Numerator}).
This condition induces the following relations:
\begin{align}
&|\tilde{\Psi}_{0}^{(0)}(\vec{k}_{1}+\Delta\vec{k})|^{2}\prod_{i\neq 0}(E_{0}+\delta-E_{i})+\notag \\
&\delta\sum_{j\neq0}|\tilde{\Psi}_{j}^{(0)}(\vec{k}_{1}+\Delta\vec{k})|^{2}\prod_{i\neq 0,j}(E_{0}+\delta-E_{i})=0,\\
&|\tilde{\Psi}_{j}^{(0)}(\vec{k}_{1}+\Delta\vec{k})|
=|\vec{e}_{j}(\vec{k}_{0})^{\dagger}\cdot\vec{e}_{0}(\vec{k}_{1}+\Delta\vec{k})|\notag\\
&\sim |\tilde{\Psi}_{j}^{(0)}(\vec{k}_{1})|+O(\Delta\vec{k})~~\text{for $j\neq0$}.
\end{align}
By taking the lowest order of $\delta$ and $\Delta k$,
we obtain
\begin{align}
&|\tilde{\Psi}_{0}^{(0)}(\vec{k}_{1}+\Delta\vec{k})|^2\sim \delta\times\Big[\frac{K(\vec{k}_{1})}{A(\vec{k}_{1})}\Big] \\
&K(\vec{k}_{1})=-\sum_{j\neq0}|\tilde{\Psi}_{j}^{(0)}|^2B_{j}(\vec{k}_{1})\\
&B_{j}(\vec{k}_{1})=\prod_{i\neq 0,j}[E_{0}(\vec{k}_{1})-E_{i}(\vec{k}_{1})] \\
&A(\vec{k}_{1})=\prod_{i\neq 0}[E_{0}(\vec{k}_{1})-E_{i}(\vec{k}_{1})]
\end{align}
We note that $\delta>0$ indicates that $\tilde{\zeta}_{0}$
traverse the band gap. Therefore the condition of the
traverse is given by $[K(\vec{k}_{1})/A(\vec{k}_{1})]\geq0$.

We obtained the following relation:
{\begin{align}
\frac{K(\vec{k}_{1})}{A(\vec{k}_{1})}=
-\sum_{j<0}\frac{|\tilde{\Psi}_{j}^{(0)}|^2}{|E_{0}-E_{j}|}
+\sum_{j>0}\frac{|\tilde{\Psi}_{j}^{(0)}|^2}{|E_{0}-E_{j}|}
\label{eq:inversion}
\end{align}}
Therefore, when $\vec{e}_{0}(\vec{k}_{1})$ 
consists of only the unoccupied eigenvectors of $\vec{k}_{0}$,
$K(\vec{k}_{1})/A(\vec{k}_{1})$ is definitively positive 
since $\tilde{\Psi}_{j}^{(0)}=0$ for $j<0$.
This means $\delta$ that
$\tilde{\zeta}_{0}$ traverse the band gap.
Even when $\vec{e}_{0}(\vec{k}_{1})$ contains 
both the unoccupied and occupied eigenvectors, $K(\vec{k}_{1})/A(\vec{k}_{1})$
can be positive if the weight of the unoccupied bands is dominant.

For $K(\vec{k}_{1})=0$, the 1st order of $\delta$ vanishes. 
By considering the 2nd order of $\delta$,
we obtain the relation as
{\begin{align}
&|\Psi_{0}^{(0)}(\vec{k}_{1}+\Delta\vec{k})|^2\sim \delta^{2}
\times\Big[\frac{C(\vec{k}_{1})}{A(\vec{k}_{1})}\Big],\\
&\frac{C(\vec{k}_{1})}{A(\vec{k}_{1})}=\sum_{j\neq0}
\frac{|\Psi_{j}^{(0)}|^2}{|E_{0}-E_{j}|^2}
\end{align}}
Since $C(\vec{k}_{1})/A(\vec{k}_{1})$ is always positive,
$\delta$ has both the positive and the negative solutions.
This indicates that both zeros below and above $E_{0}$
coincide at $\vec{k}_{1}$ for $K(\vec{k}_{1})=0$.
Therefore, even for $K(\vec{k}_{1})=0$,
$\tilde{\zeta}_{0}$ traverses the band gap.

The same argument can be applied to the lowest unoccupied bands.
When $\vec{e}_{1}(\vec{k}_{1})$ contains the occupied eigenvectors of $\vec{k}_{0}$ 
and their weight is dominant,
$\tilde{\zeta}_{1}$ traverse the band gap and the crosses of the zeros appear.

\section{AI topological insulator in four dimensions}
\label{ap:AI}

\begin{figure}[tb]
	\begin{center}
		\includegraphics[width=8cm]{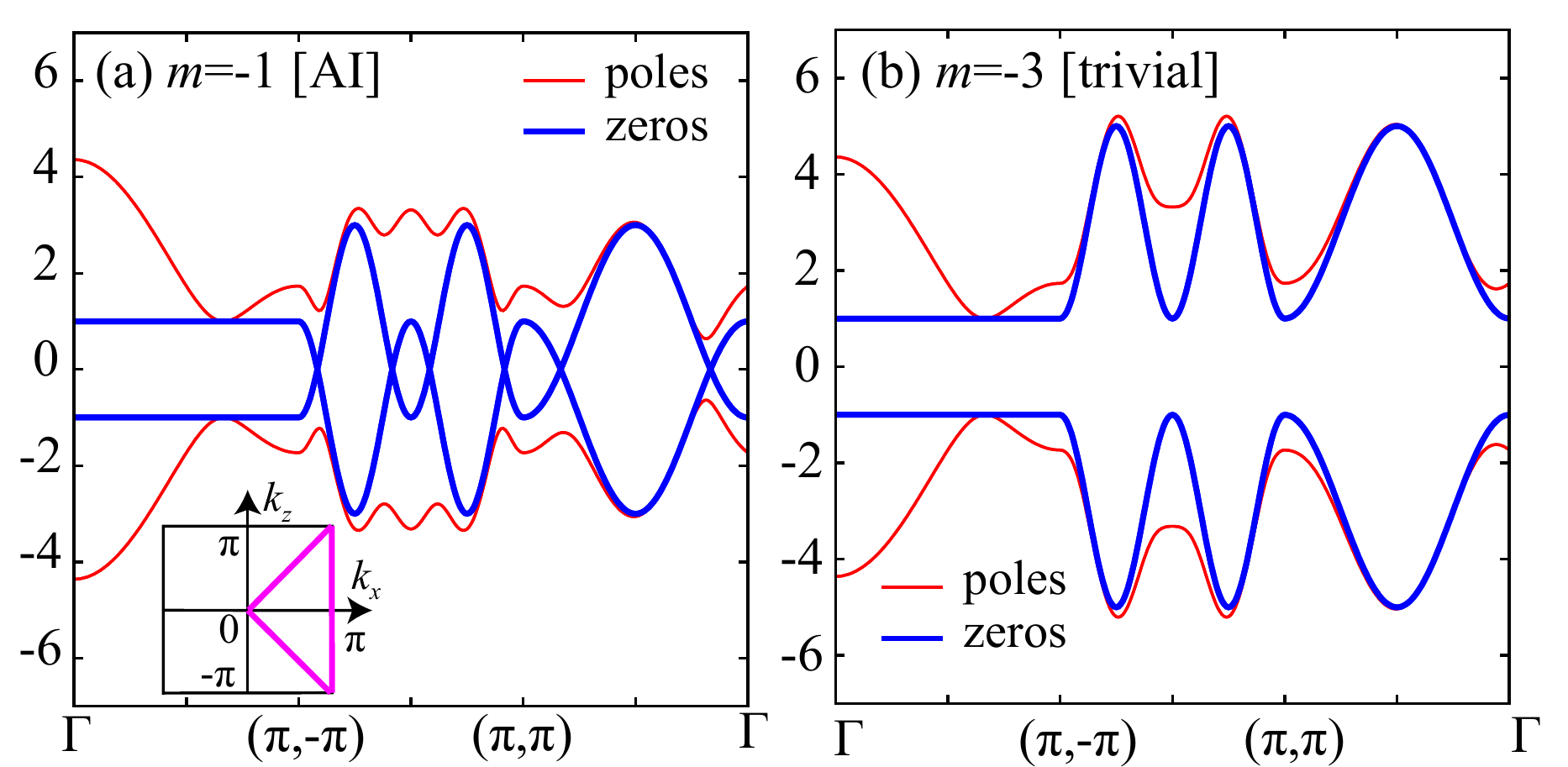}   
	\end{center}
\caption{~Band dispersions and zeros of the Green functions
in the model defined in Eq.~(\ref{eq:AI}) for the (a) topological phase
and (b) the trivial phase.
For this study, $J^{\prime}=1,J^{\prime\prime}=0$, and $k_y=k_w=0$.
For $-2J^{\prime}<m<J^{\prime}$ 
($J^{\prime\prime}=0$), the system becomes a topological insulator.
In the inset, the symmetry line that was used in the plot is illustrated.
The zeros {traverse the bandgap} in 
the AI topological insulators.
}
\label{fig:AI}
\end{figure}%

According to the periodic table for the topological insulators,
the AI topological insulator does not appear in the
realistic dimensions ($d=1,2,3$), but it appears
in four dimensions.
The model Hamiltonian is proposed in Ref.~[\onlinecite{Price_PRB2020}].
Its realization in the electric circuits has been proposed~\cite{Wang_NCom_2020}.
The Hamiltonian for the AI topological insulators is given as
\begin{align}
&H= \notag \\
&\begin{pmatrix}
M              & ~0             & ~R_{z}-iA     & R_{x}-iR_{y} \\
0              & ~M             & ~R_{x}+iR_{y} & ~-R_{z}-iA   \\
R_{z}+iA       & ~R_{x}-iR_{y}  & ~-M           & ~0            \\
R_{x}+iR_{y}   & ~-R_{z}+iA     & ~0            & ~-M           \\
\end{pmatrix},
\label{eq:AI}
\end{align}
where $R_{x}=2\cos{k_z}+\cos{k_w}$,
$R_{y}=\sin{k_w}$, $R_{z}=2\cos{k_x}+\cos{k_y}$,
$A=\sin{k_y}$, and $M=m+2J^{\prime}\cos{(2k_x+2k_z)}+2J^{\prime\prime}\cos{(2k_x-2k_z)}$. 
It can be noted that $k_{w}$ is the momentum in the fourth dimension.

The eigenvalues (doubly degenerate) are given by
\begin{align}
E_{0} &= \sqrt{M^2+A^2+R_{x}^2+R_{y}^2+R_{z}^2}, \\
E_{1} &= -\sqrt{M^2+A^2+R_{x}^2+R_{y}^2+R_{z}^2},
\end{align}
and the zeros of the Green functions are
given by
\begin{align}
\zeta_{0} &= -M \notag \\ 
          &= -(m+2J^{\prime}\cos{(2k_x+2k_z)}+2J^{\prime\prime}\cos{(2k_x-2k_z)}), \\
\zeta_{1} &= M  \notag \\
          &= m+2J^{\prime}\cos{(2k_x+2k_z)}+2J^{\prime\prime}\cos{(2k_x-2k_z)}.
\end{align}
For $J^{\prime\prime}=0$,
it is shown that
the system becomes an AI topological insulator 
for $-2J^{\prime}<m<J^{\prime}$~~\cite{Price_PRB2020}.
As shown in Fig.~\ref{fig:AI},
in the AI topological insulator,
the zeros of the Green functions {traverse} the bandgap.

\section{Proof of the relation with the $m$-fold degenerate edge states and
the zeros in $\bar{G}$ I}
\label{appendixB}
The bulk Hamiltonian for the $m$-orbital system is given as
\begin{align}
H_{\rm real}=
\begin{pmatrix}
A           ~~~& B       \\
B^{\dagger} ~~~& H_{\rm edge}
\end{pmatrix},
\label{eq:Hreal}
\end{align}
where the size of $A$ is $m\times m$,
the size of $B$ is $n\times m$,
and the size of $H_{\rm edge}$ is $n\times n$.
As we explained in Sec.~\ref{sec:zeroedge}, 
$H_{\rm edge}$ describes the system with the edges or surfaces.
It is assumed that $n$ is sufficiently larger than $m$.
By using the unitary matrix that diagonalizes
$H_{\rm edge}$, the Hamiltonian can be transformed as
\begin{align}
&\tilde{H}_{\rm real}=
\begin{pmatrix}
I          ~~~& 0           \\
0          ~~~& U^{\dagger} 
\end{pmatrix}
\begin{pmatrix}
A           ~~~& B       \\
B^{\dagger} ~~~& H_{\rm edge}
\end{pmatrix}
\begin{pmatrix}
I          ~~~& 0           \\
0          ~~~& U
\end{pmatrix} \\
&=
\begin{pmatrix}
A                      ~~~ & \tilde{B} \\
\tilde{B}^{\dagger}    ~~~ & D
\end{pmatrix} \\
&D={\rm diag}(E_{0},E_{1},\ldots,E_{n-1}),
\end{align}
where $E_{i}$ represents the $i$th eigenvalue of $H_{\rm edge}$.

Here, it is assumed that $H_{\rm edge}$ has $m$ degenerate eigenvalues.
Without loss of generality,
it can be assumed that $E=E_{0}=E_{1}=\cdots=E_{m-1}$.
The following form of the inverse of $\bar{G}$ can be obtained.
\begin{align}
&\tilde{\bar{G}}_(\omega=E)^{-1}=(EI-\tilde{H}_{\rm real})= \notag\\
&\begin{pmatrix} 
G_{A}(E)^{-1}       ~~~ & \tilde{F}      ~~~& \tilde{C}       \\
\tilde{F}^{\dagger} ~~~ & 0_{m\times m}  ~~~& 0   \\
\tilde{C}^{\dagger} ~~~ & 0              ~~~& \tilde{G}_{D_{n-m}}(E)^{-1}  \\
\end{pmatrix},
\end{align}
where 
$G_{A}(E)^{-1}=(EI-A)$,$\tilde{G}_{D_{n-m}}(E)^{-1}=(EI-D_{n-m})$, and
$D_{n-m}={\rm diag}(E_{m},\ldots,E_{n-1})$.
From this, the bulk Green function can be expressed as
\begin{align}
&\tilde{\bar{G}}(\omega=E)= \notag \\
&\begin{pmatrix} 
0_{m\times m}  ~~~ & (\tilde{F}^{\dagger})^{-1}                           ~~~& 0                                          \\
\tilde{F}^{-1} ~~~ & X                                                    ~~~& -\tilde{F}^{-1}\tilde{C}\tilde{G}_{D_{n-m}}   \\
0              ~~~ & -\tilde{G}_{D_{n-m}}\tilde{C}^{\dagger}(\tilde{F}^{\dagger})^{-1}  ~~~& \tilde{G}_{D_{n-m}},                          
\end{pmatrix}, \\
\label{eq:invG}
&X=-\tilde{F}^{-1} G_{A}^{-1}(\tilde{F}^{\dagger})^{-1} +\tilde{F}^{-1}\tilde{C}\tilde{G}_{D_{n-m}}\tilde{C}^{\dagger}(\tilde{F}^{\dagger})^{-1},
\end{align}
{where we assume that $\tilde{F}^{-1}$ exists.}

$\tilde{\bar{G}}_{\rm bulk}(\omega=E)$ can be rewritten as
\begin{align}
\tilde{\bar{G}}_{\rm bulk}(\omega=E) =
\begin{pmatrix}
0_{m\times m} ~~~& b           \\
b^{\dagger}   ~~~& \tilde{X}
\end{pmatrix}.
\end{align}
By using the relation
\begin{align}
&\bar{G}(E) = (EI-H_{\rm real})^{-1}
= \tilde{U}\tilde{\bar{G}}_{\rm real}\tilde{U}^{\dagger}, \\
&\tilde{U}=
\begin{pmatrix}
I          ~~~& 0           \\
0          ~~~& U
\end{pmatrix},
\end{align}
we can obtain
\begin{align}
\bar{G} = 
\begin{pmatrix}
0_{m\times m} ~~~& bU^{\dagger}           \\
Ub^{\dagger}  ~~~& U\tilde{X}U^{\dagger}
\end{pmatrix}.
\end{align}
Therefore, the zeros of $\bar{G}$
have $m$ degeneracies when $H_{\rm edge}$
has $m$ degenerate eigenvalues.
{We note that this argument can not be applied for $m_{\rm edge}<m_{\rm orb}$
since Eq.~(\ref{eq:invG}) does not hold.}

\section{Proof of the relation with the $m$-fold degenerate edge states and
the zeros in $\bar{G}$ II}
\label{appendixC}
We show another proof of the relation using the Cauchy interlacing identity.
Here, for $H_{\rm real}$ defined in Eq.~(\ref{eq:Hreal}), 
we define $k$th sub matrices $M^{k}(i_{0},i_{1},\cdots,i_{k-1})$ 
that remove $k$ columns and rows from the original Hamiltonian $H_{\rm bulk}$.
The indices of the removal columns and rows are represented 
by $i_{0},i_{1},\cdots,i_{k-1}$ 
($i_{\alpha}\neq i_{\beta}$ for $\alpha\neq\beta$). 
We can take $0\leq i_{\alpha}\leq m-1$ without loss of generality from the periodicity of $H_{\rm bulk}$.
The $i$th ($i=0,1,\cdots,m-1$) zeros of $\bar{G}$ ($\bar{\zeta}_{i}$) are
given by the eigenvalues of the first minor matrix $M^{1}(i)$. 

\begin{figure}[h!]
	\begin{center}
	\includegraphics[width=5cm,clip]{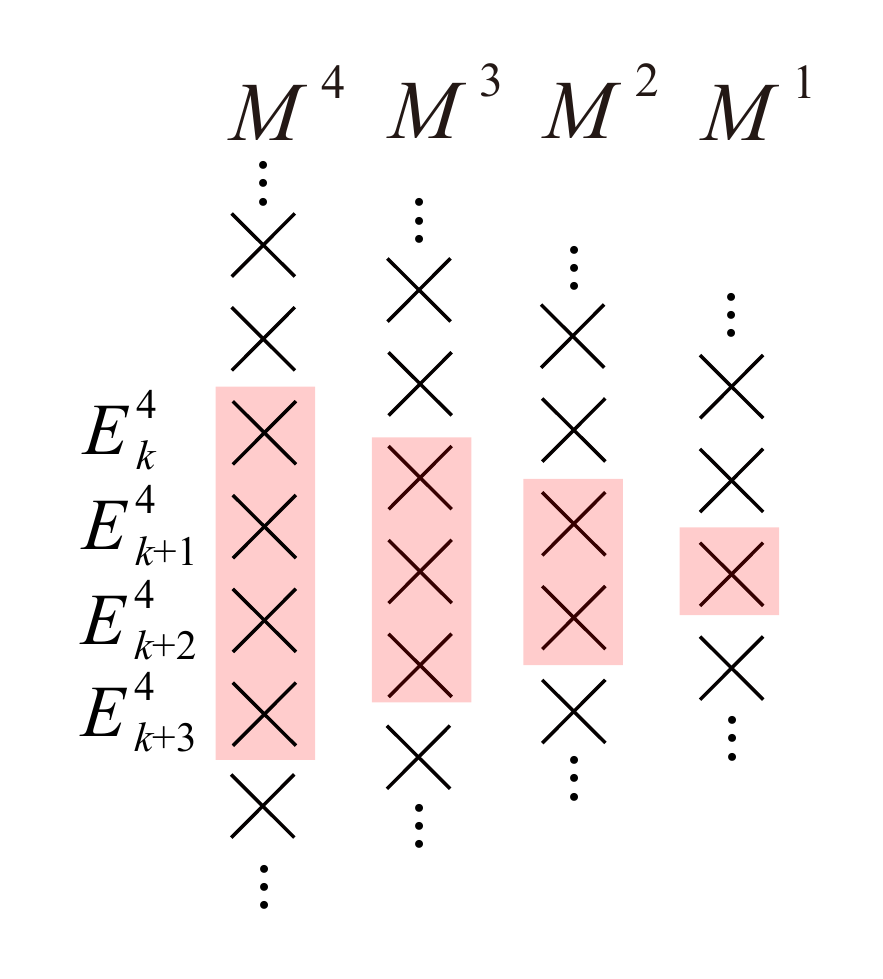}   
	\end{center}
\caption{Schematic picture of the eigenvalues (denoted by crosses) of 
the sub matrices for 4-orbital system. 
From the Cauchy interlacing inequality, the
eigenvalues of $M^{k-1}$ should exist between the
eigenvalues of $M^{k}$.
If the eigenvalues of $M^{4}$ have 4-fold degeneracy
($E=E^{4}_{k}=E^{4}_{k+1}=E^{4}_{k+2}=E^{4}_{k+3}$),
one of the eigenvalues of $M^{1}$ should coincide
with $E$. The shaded crosses have the same energy $E$.
This indicates the zeros have the $4$-fold degeneracy.}
\label{fig:ZEC}
\end{figure}

If the eigenvalues of $M^{m}(0,1,\cdots,m-1)=H_{\rm edge}$
have $m$-fold degeneracy ($E=E^{m}_{k}=E^{m}_{k+1}=\cdots=E^{m}_{k+m-1}$), 
the eigenvalues of $M^{m-1}$ have $(m-1)$-fold degeneracy
($E=E^{m-1}_{k}=E^{m-1}_{k+1}=\cdots=E^{m-1}_{k+m-2}$)
from the Cauchy interlacing inequality.
By using the Cauchy interlacing inequality iteratively,
we can show that one of the eigenvalues of 
the 1st minor matrix $M^{1}(i)(i=0,1,\cdots m-1)$ 
is the same as $E$.
In Fig.~\ref{fig:ZEC}, we show the schematic representation of this 
relation for the 4-orbital systems.
This indicates that the zeros of the Green functions have 
$m$-fold degeneracy, i.e., $E=\bar{\zeta_{0}}=\bar{\zeta}_{1}=\cdots=\bar{\zeta}_{m-1}$.
{Obviously, this argument can not be applied for $m_{\rm edge}<m_{\rm orb}$
since the Cauchy interlacing inequality can be used at most $m_{\rm edge}$ times.}

\providecommand{\noopsort}[1]{}\providecommand{\singleletter}[1]{#1}%
\end{document}